\documentclass[twoside]{elsart}
\usepackage{graphicx}
\def\Journal#1#2#3#4{{#1} {\bf #2}, #3 (#4)}

\def\NIMA{{\em Nucl. Instrum. Methods} A}

\def\PLB{{\em Phys. Lett.}  B}

\def\kos{K^0_S}

\begin{document}                                                 

\def\bibname{References}
\bibliographystyle{plain}

\begin{frontmatter}

\title{RECONSTRUCTION OF VEES, KINKS, $\Xi^-$'s, and $\Omega^-$'s
IN THE FOCUS SPECTROMETER}
%

\long\def\inst#1{\par\nobreak\kern 4pt\nobreak
    {\it #1}\par\vskip 10pt plus 3pt minus 3pt}
The FOCUS Collaboration

\author[ucd]{J.~M.~Link},
\author[ucd]{M.~Reyes\thanksref{atmichoacana}},
\author[ucd]{P.~M.~Yager},
\author[cbpf]{J.~C.~Anjos},
\author[cbpf]{I.~Bediaga},
\author[cbpf]{C.~G\"obel\thanksref{aturuguay}},
\author[cbpf]{J.~Magnin},
\author[cbpf]{A.~Massafferri},
\author[cbpf]{J.~M.~de~Miranda},
\author[cbpf]{I.~M.~Pepe\thanksref{atbahia}},
% ...

\address[ucd]{University of California, Davis, CA 95616} 
\address[cbpf]{Centro Brasileiro de Pesquisas F\'isicas, Rio de Janeiro, RJ, Brasil} 
\address[cinv]{CINVESTAV, 07000 M\'exico City, DF, Mexico} 
\address[cu]{University of Colorado, Boulder, CO 80309} 
\address[fnal]{Fermi National Accelerator Laboratory, Batavia, IL 60510} 
\address[fras]{Laboratori Nazionali di Frascati dell'INFN, Frascati, Italy I-00044} 
\address[ui]{University of Illinois, Urbana-Champaign, IL 61801} 
\address[iu]{Indiana University, Bloomington, IN 47405} 
\address[korea]{Korea University, Seoul, Korea 136-701} 
\address[milan]{INFN and University of Milano, Milano, Italy} 
\address[nc]{University of North Carolina, Asheville, NC 28804} 
\address[pavia]{Dipartimento di Fisica Nucleare e Teorica and INFN, Pavia, Italy} 
\address[pr]{University of Puerto Rico, Mayaguez, PR 00681} 
\address[sc]{University of South Carolina, Columbia, SC 29208} 
\address[ut]{University of Tennessee, Knoxville, TN 37996} 
\address[vu]{Vanderbilt University, Nashville, TN 37235} 
\address[wisc]{University of Wisconsin, Madison, WI 53706} 

\thanks[atmichoacana]{Present Address: Instituto de Fisica y Matematicas, Universidad Michoacana de San Nicolas de Hidalgo, Morelia, Mich., Mexico 58040} 
\thanks[aturuguay]{Present Address: Instituto de F\'isica, Faculdad de Ingenier\'i a, Univ. de la Rep\'ublica, Montevideo, Uruguay} 
\thanks[atbahia]{Present Address: Instituto de F\'isica, Universidade Federal da Bahia, Salvador, Brazil} 
\thanks[atmilan]{Present Address: INFN sezione di Milano, Milano, Italy} 
\thanks[atauton]{Present Address: Instituto de Ciencias Nucleares, Universidad Nacional Aut\'onoma de M\'exico. CP 04510. M\'exico}
\thanks[atpueblo]{Present Address: Instituto de F\'{\i}sica, Universidad Aut\'onoma de Puebla, Puebla, M\'exico} 
\thanks[atncar]{Present Address: National Center for Atmospheric Research, Boulder, CO} 
\thanks[atvu]{Present Address: Vanderbilt University, Nashville, TN 37235} 
\thanks[intel]{Present Address: Intel Corporation, Portland Technology Development, RA1-238 5200 N.E. Elam Young Parkway Hillsboro, OR 97124}
\thanks[roches]{Present Address: University of Rochester, Fermilab, P.O. Box 500, Batavia, IL 60510}
\thanks[atkriss]{Present Address:  Korea Research Institute of Standards and Science, P.O.Box 102, Yusong-Ku, Taejon 305-600, South Korea} 
\thanks[atkyung]{Present Address: Center for High Energy Physics, Kyungpook National University, 1370 Sankyok-dong, Puk-ku, Taegu, 702-701, Korea}
\thanks[atcu]{Present Address: University of Colorado, Boulder 80309} 
\thanks[atbrescia]{Present Address: Dipartimento di Chimica e Fisica per l'Ingegneria e per i Materiali, Universit\'a di Brescia and INFN sezione di Pavia} 
\thanks[atbucharest]{Present Address: Nat. Inst. of Phys and Nucl. Eng., Bucharest, Romania} 
\thanks[atlucent]{Present Address: Lucent Technology} 
\thanks[atadamst]{Present Address: Adams State College, Alamosa, CO 81102}

%

\begin{abstract}
We describe the various techniques developed
in the Fermilab Wideband Experiments, E687 and FOCUS, to
reconstruct long-lived states. The techniques all involve
modifications to standard tracking techniques and are
useful to report for future experiments.
\end{abstract}

\end{frontmatter}

\section{\bf Introduction}
FOCUS is a photoproduction experiment located at Fermilab which has
been configured to
investigate the production
and decay of
charmed particles. As charmed particles typically decay to states
containing strange particles, techniques have been developed to
reconstruct long-lived mesons and baryons containing strange quarks.
The techniques employ a full range of information from a
multipurpose spectrometer including partially reconstructed tracks
and kinematic
constraints,  and utilize excellent charged particle identification using \v{C}erenkov
detectors, efficient muon identification,
and the ability to reconstruct neutral hadronic energy.

In this paper we will briefly describe the spectrometer in Section 2.
In Section 3 we will discuss how the tracking is performed in the
silicon microstrip detector and in the multiwire proportional 
chambers and how the
individual segments are linked to form tracks. 
In Section 4 we describe our
neutral Vee reconstruction techniques. 
In Section 5 we
present our algorithms for identifying $\Sigma$ decays where one
decay product goes undetected (Kinks).
In Section 6 we present our
methods for combining
of the Vee and Kink algorithms to find $\Xi^-$ and $\Omega^-$ decays.

\section{\bf Spectrometer Components and Layout}
\begin{figure*}[hbtp]
\begin{center}
{\includegraphics[width=14.0cm]{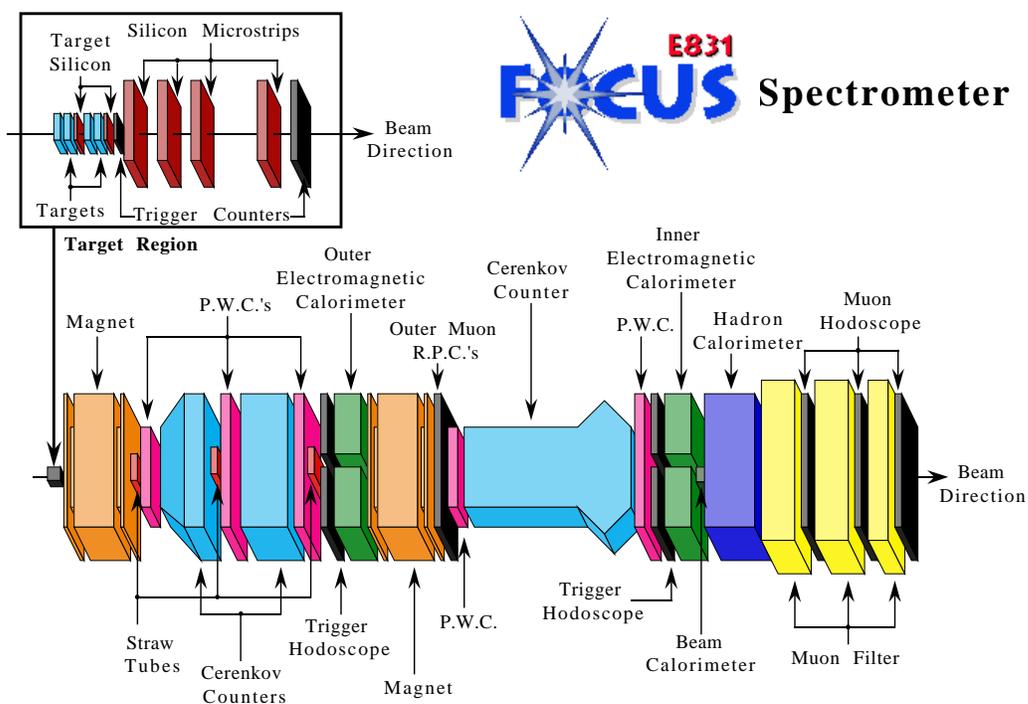}}
\end{center}
\caption{A schematic drawing of the FOCUS spectrometer. The inset
displays the segmented targets, the embedded target silicon, 
the trigger counters, and the 12 plane silicon tracking array.
The spectrometer is about 30 meters long.}
\label{fig:spect}
\end{figure*}

The spectrometer shown in Fig. \ref{fig:spect}
consists of two dipole magnets, M1 and M2; a 12 plane silicon
microstrip array; five stations of multiwire proportional
chambers;
300 threshold
\v{C}erenkov cells arranged in three gas boxes; 
two electromagnetic calorimeters for photon, electron, and
$\pi^0$ reconstruction; two muon detectors; and a hadronic
calorimeter for triggering and neutral hadron reconstruction.
There are 10 meters along the spectrometer in which Vees can
can be reconstructed. There is also a large 
charged particle acceptance covering the entire forward
hemisphere. The FOCUS spectrometer is an upgraded version
of the previous experiment E687. A detailed description of the 
E687 spectrometer
and its performance may be found in Reference~1.

The two large aperture magnets are used 
to momentum analyze charged tracks. The magnets are oriented
to bend particles in the vertical ({\it y}-axis) direction with 
M1 and M2 bending in {\it opposite} directions. Positive particles
are bent upwards in M1. The particular
arrangement of magnet positions and orientation was chosen for
its unique effect on event topology. There is a large background
of ~$e^+e^-$ pairs coming from beam photon conversions in the 
experimental target. Pairs are produced with little
transverse momentum with respect to the 
beam direction ({\it z} direction)
and with a transverse profile comparable to the beam size (about
1~cm in {\it x} and {\it y}). The first magnet bends the electrons and
positrons in {\it y}, creating a vertical swath. The lowest energy
pairs strike the inside of M1 or the upstream face of M2,
while the remainder pass through the M2 aperture and are bent 
back towards the beam axis. The beam profile is reconstituted
at the end of the spectrometer, with some smearing due to 
energy loss  via bremsstrahlung from material
throughout the spectrometer.

The target is composed of segmented
BeO slabs with silicon microstrip doublets after the first
two targets and after the last two targets. For historical 
reasons these four
silicon microstrip planes are not used directly in the track
reconstruction, however they are used to extend the microstrip
tracks to the production and secondary vertices. By making
a measurement close to the vertex we are able to 
improve both the position and  the angular resolution of the
track.

\section {\bf Standard Tracking Devices and Algorithms}
\subsection {\bf Microstrip Detector}
The high resolution tracking of charged
particles provided by the Silicon microStrip Detector, or the
SSD, is essential to the reconstruction
of charmed particle decay vertices and in separating these vertices from
the production vertex.   

\begin{figure*}[htbp]
\begin{center}
{\includegraphics[width=14cm]{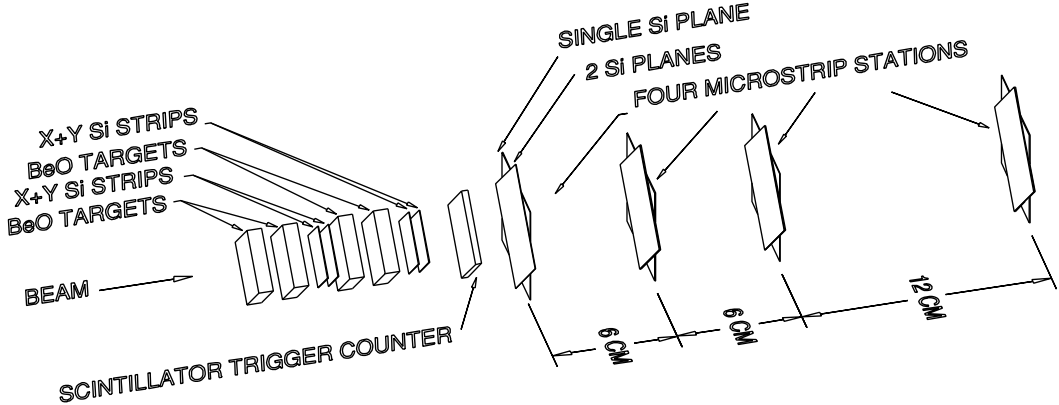}}
\end{center}
\caption{A layout drawing of the silicon microstrip detector
together with the BeO production targets and embedded 
target silicon strip planes.} 
\label{fig:micro}
\end{figure*}

The SSD consists of four physically separated
stations with three planes per station. The three planes of 
each station are oriented to {\it i, j,} and {\it k} coordinates
of -135, -45, and -90 degrees with respect to the horizontal ({\it
x} axis). The microstrip detector and target layout is shown in
Fig. \ref{fig:micro} and a detailed summary of the active 
area, high resolution region, strip pitch, and number of channels
for each station is given in Table  \ref{tab:ssdtab}.

The innermost section of each plane, covering the region
where tracks pass most closely to each other, has twice the 
resolution of the outer section. In addition, the 
most upstream station has a resolution  twice as good as the 
other stations. Pulse heights are read out for all 8,256
strips. The overall detection efficiency of each plane is better
than 99\% even including the non-functioning strips and broken electronics
channels. A more complete description of this detector
and its performance can be found in Reference~2.

\begin{table*}[htbp]
\caption{Properties of the Silicon Microstrip Detector. The first 
three lines are the internal relative z locations of the silicon
microstrip planes.
\label{tab:ssdtab}}
\vspace{0.4cm}
\begin{center}
\begin{tabular}{lllll}
\hline
 Property  & I station & II station & III station & IV station \\
\hline
 $1^{st}$~detector & -0.5 {\textrm cm} & 5.5 {\textrm cm} & 
11.5 {\textrm cm} & 23.5 {\textrm cm} \\
 $2^{nd}$~detector & 0.0~cm & 6.0~cm & 12.0~cm & 24.0~cm \\
 $3^{rd}$~detector & 0.5~cm & 6.5~cm & 12.5~cm & 24.5~cm \\
 active area & 2.5 $\times$ 3.5~cm$^2$ & 5 $\times$ 5~cm$^2$
& 5 $\times$ 5~cm$^2$ & 5 $\times$ 5~cm$^2$ \\
 high res area & 1.0 $\times$ 3.5~cm$^2$ & 2 $\times$ 5~cm$^2$ & 
2 $\times$ 5~cm$^2$ & 2 $\times$ 5~cm$^2$\\
 strip pitch & 25~$\mu$m, 50~$\mu$m & 50~$\mu$m, 100~$\mu$m
& 50~$\mu$m, 100~$\mu$m & 50~$\mu$m, 100~$\mu$m \\
 \# of channels & 688 $\times$ 3 &  688 $\times$ 3 &  688 $\times$ 3&
688 $\times$ 3 \\
\hline
\end{tabular}
\end{center}
\end{table*}

The microstrip tracking algorithm is based on projection
finding using hits in three separate views.
A hit is a cluster of one to three adjacent 
strips depending on the summed pulse heights in the strips 
(less than a 1.5 minimum
ionizing pulse height) and on the number of neighboring strips
with pulse heights above threshold.
As momentum information
is not yet available multiple Coulomb scattering effects
are not considered. Projections are found using very loose
selection criteria.
At least three hits per view (out of a possible four)
are required and sharing of hits is permitted. Hits in
the last three stations that are already assigned to a projection
having hits in all four stations cannot be reused for a new 
projection containing only three hits. Projections are combined
to form tracks if they match in space with a global $\chi^2$
per degree of freedom less than 8. Tracks sharing projections
are arbitrated according to their $\chi^2$. The process is
performed in a fully symmetric way with respect to the views.

The pulse height information on each hit is used in three ways.
First, the information allows for the separation of two overlapping
hits in the pattern recognition. Second, the pulse height 
information is combined from all hits in the SSD in the track to determine 
whether the track is consistent with the passage of a singly 
charged track through the planes. By requiring more than 1.5~times
a standard singly charged pulse height we can identify $e^+e^-$ 
pairs. Third, and most importantly, we use the charge sharing
between neighboring strips to obtain an improved position 
resolution.

\subsection {\bf Multiwire Proportional Chambers (MWPCs)}

Five stations of Multiwire Proportional Chambers are used
to track charged particles in the main spectrometer. The
first three chamber stations, labelled P0, P1, and P2, 
are located between
the two analyzing magnets, M1 and M2. The other two chamber
stations, P3 and P4, are located downstream of the second
magnet, M2. This arrangement allows for two momentum 
measurements for higher energy tracks which pass through 
both magnets 
as well as providing momentum information
for wide angle tracks which are not accepted by the M2 aperture.

Each of the chamber stations consist of four planes of wires,
measuring {\it y}, {\it u}, {\it v}, and {\it x} positions. 
The X-view wires, running
vertically, measure the horizontal (non-bend view) position.
The U-view and V-view wires are oriented at $\pm$~11.3 
degrees with respect
to the Y-view, an arrangement which is used to resolve ambiguities
and to provide better momentum resolution.

The tracking algorithm for the multiwire proportional chambers
is used to reconstruct spectrometer tracks which have the first hit 
in P0 (the
most upstream chamber) and which extend into at least one additional 
chamber. As with microstrip tracks, a projection finding 
algorithm is used. The X-view is special because it is
the only MWPC view in which the charged particles are not
bent.

First a microstrip track is extended into the
non-bend X-view of the chambers. A search is made for MWPC hits
which match this ``seed" projection. Next, projections are
formed from hits in the U-view, V-view, and Y-view. The projections in
all four views are combined to form tracks. Further X-view projections
are formed using unused MWPC hits where no seed track exists and these
projections are also matched with unused 
U-view, V-view, and Y-view projections to reconstruct
additional tracks. Tracks are required to extend through at least
the first three MWPCs.  A least squares fit is performed 
on all candidate
tracks. The fit parameters are the intercepts  and slopes of 
the track in the {\it xz} and {\it yz} planes. If a track
passes through just the first three MWPCs, 
then it is referred to as a {\it stub}. For tracks passing through
all five chamber stations (called 5-chamber tracks or just {\it tracks}),
the change in the {\it y} slope between the track segments upstream and
downstream of M2, {\it i.e.} the bend angle, is an additional 
parameter.   

An approximate momentum is assigned to 5-chamber tracks by using
the bend angle in M2 and the sudden bend approximation. Various
magnetic corrections are applied to this category of tracks.
Fringe fields in M1 and M2, extending beyond the magnetic poles,
and off-field components  of the magnetic field ({$B_y$ and $B_z$)
are treated by higher order corrections to the linear fit. The
corrections are applied by iterating each 5-chamber track through
a complete fit which includes the magnetic corrections. The track
momentum is adjusted appropriately after each iteration. 

\subsection {\bf Linking}

Once tracks are reconstructed in the silicon microstrip and MWPC
systems, they must be linked together to determine which microstrip track
is associated to which MWPC track. This is accomplished by
extrapolating each microstrip track to the center of M1 and 
searching for MWPC tracks which, when extrapolated back to this
same point in {\it z}, ``match" with it. This matching is performed 
by comparing
the MWPC track's extrapolated {\it x} slope and {\it x} intercept at the
center of M1 with the same quantities for the microstrip track
in question. Because the magnetic field is in the {\it x} direction 
there is no (or little)  bending of charged tracks in this view. 
Thus, for a MWPC track and a microstrip track created by the 
same particle these two {\it x} quantities should agree. (This match
is accomplished after a correction is made for weak focusing in the 
non-bend view.)
 A global least 
squares fit using all the hit information from the microstrip 
segment and the 
MWPC segment is then performed on each candidate. Multiple 
MWPC segments linked to the same microstrip segment are 
arbitrated on the basis of the $\chi^2$ per degree of freedom
of the fit. A maximum of two MWPC tracks are allowed to be
linked to the same microstrip track. This decision is made
because $e^+e^-$
pairs from beam photon conversions frequently reconstruct as
a single microstrip track due to their extremely small opening
angles, but will reconstruct as two separate MWPC tracks as
the $e^+$ and $e^-$ are bent in opposite directions in M1.  

Track segments from the microstrip tracks and the MWPC tracks
which fail to link are used to form Vees, Kinks, $\Xi^-$'s, 
and $\Omega^-$'s. 

\subsection{\bf Primary Vertex}

In the early stages of data reconstruction a basic vertex algorithm
is implemented using only the microstrip track information. 
These vertices are used for reconstructing different Vee categories
and are meant to be roughly correct. Later in our analyses the ``true" 
primary vertex is recalculated using all available tracking information
with multiple scattering information included. The exact algorithm
depends on the 
final state that is
being investigated. For instance we use a candidate driven approach
when the decay daughters are fully reconstructed such as 
$D^0\rightarrow K^-\pi^+$, a special seed plane approach for one 
prong final states such as $D^+\rightarrow K^0_S\pi^+$, and 
a stand alone vertexing algorithm for semileptonic decays such
as $D^+\rightarrow K^-\pi^+\mu^+\nu$. These techniques have been
described in several E687 theses. A good reference is the 
thesis~\cite{cinquini} of Luca Cinquini. 

The basic vertex algorithm minimizes the distance of closest
approach of the tracks in the transverse plane. Specifically,
the minimized quantity is:

\begin{equation}
\chi^2 = \sum_{i=1}^n \biggl({{X-(x_i + a^{\prime}_iZ)}\over {\sigma_{x,i}}}\biggr)^2
+ \biggl({{Y-(y_i + b^{\prime}_iZ)}\over {\sigma_{y,i}}}\biggr)^2 
\end{equation}

\noindent where X, Y, Z are the coordinates of the vertex taken from the
parameters of the fit, $a^{\prime}_i, b^{\prime}_i, x_i, y_i$ are
the slopes and intercepts of the i-th track, and $\sigma_{x,i},
\sigma_{y,i}$ are the errors returned by the purely geometric track
fit to the hits of the i-th track.

The algorithm begins by assigning all microstrip tracks to a common
vertex and by computing the corresponding $\chi^2$.  Tracks are removed
from this vertex one at a time, beginning with the one that gives the largest 
contribution to the $\chi^2$. Each time the vertex is refit and the
subtraction process continues until the $\chi^2$ falls below the
threshold. Since the space location may have changed in the
process, all discarded tracks are individually tested to check whether
they originate from the found vertex and if so, they are included. 
Once the construction of the first vertex is completed, the procedure 
is repeated with the remaining set of tracks. At the end of the process
each microstrip track is assigned to just one vertex, or they might
remain unassigned. For Vee finding algorithms the most upstream
microstrip vertex within the target is called the primary vertex. If
no primary vertex is found, then the center of the target is selected
as the primary vertex.

\section{\bf Vees}

 $\kos$ and $\Lambda^0$ (usually referred to as ``Vees") are
often found among the decay products of charmed particles. In FOCUS,
these particles are reconstructed through the charged decay
modes~\cite{pdg00}:

\centerline {$\kos\rightarrow \pi^+\pi^-$(BR = 68.6\%)}
\centerline {$\Lambda^0\rightarrow p~\pi^-$(BR = 63.9\%)}      

\noindent These particles are relatively long-lived with respect
to charmed particles, and may travel several meters in the 
spectrometer before decaying. Depending on the region of decay,
they leave topologically distinct tracks, but they  must be
reconstructed with different algorithms. In all, Vees can be
reconstructed in the FOCUS spectrometer over a decay length 
of about 10 meters. A sketch of the regions where different
algorithms are employed for Vee reconstruction is presented
in Fig. \ref{fig:veegeom}. 

\begin{figure*}[htbp]
\begin{center}
{\includegraphics[width=14cm]{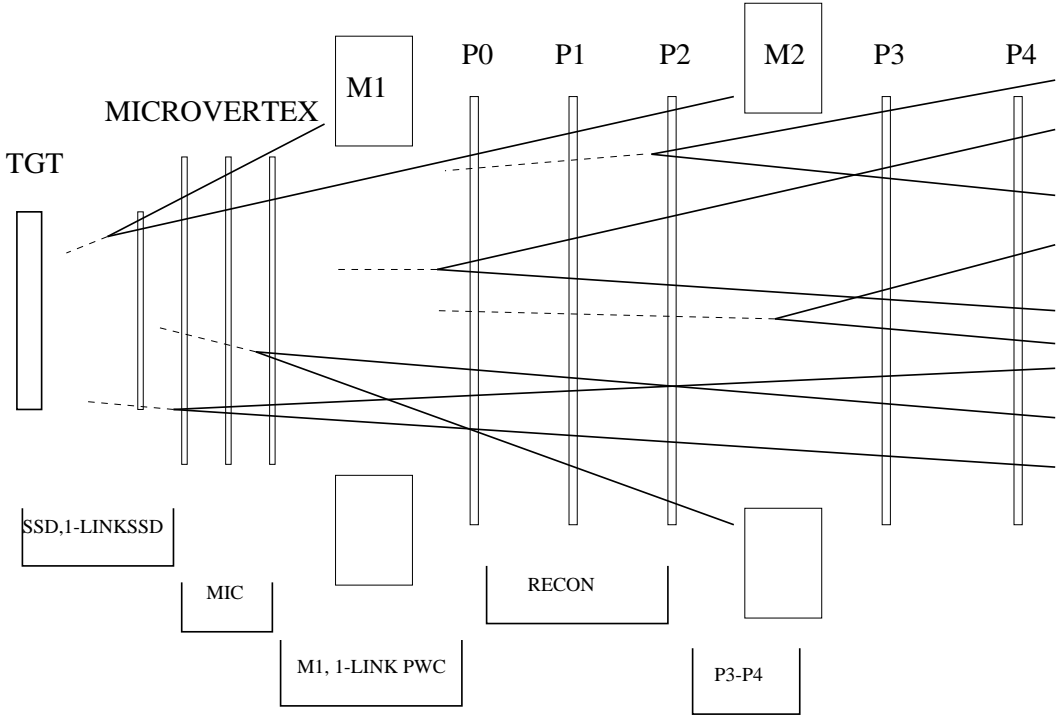}}
\end{center}
\caption{A schematic drawing of the regions of the spectrometer
in which Vees are reconstructed by different algorithms. The solid 
lines are the two-prong daughter products of the Vees. The dashed lines 
indicate that the parent particle originated from the target region.
Note that target box (TGT)  is simply for illustration. In the experiment
it is segmented and interleaved with silicon planes.}
\label{fig:veegeom}
\end{figure*}

All Vee-reconstruction codes have in
common the search for a pair of oppositely charged tracks which 
originate from a common point in space, the Vee decay vertex. The
invariant mass of the pair is computed by first assigning the pion
mass to both tracks to test the $\kos$ hypothesis. Next, the 
proton mass is assigned to the faster particle and the pion mass
to the slower particle to check for a $\Lambda^0$. Initially,
there is no \v{C}erenkov identification applied, and the Vee
requirements are intentionally left loose to allow for different
degrees of signal to noise in individual analyses.

While not all Vee categories were used in FOCUS analyses,
we did develop algorithms to identify and to reconstruct Vees 
in all regions
described in this section. The SSD Vees, M1 Vees, and One-link SSD Vees
were used in several analyses. The
RECON Vee and P34 Vee categories were used in E687, but were not
implemented in FOCUS. This resulted from a reduction in the beam energy
such that long-lived decays became relatively less important. It is
also the result of running the beam at much higher intensity which
led to increased noise in the chambers which in turn made the 
rate of false RECON Vees high and the algorithm less efficient.
For a similar reason the MIC Vee algorithm was implemented in E687,
but not in FOCUS. All MIC Vees are first found as M1 Vees and 
it required excessive computer time to search over all the 
remaining microstrip hits.  Finally, the Single-linked MPWC 
Vees were only used in searching for $\Xi^-$ and $\Omega^-$ 
decays.

\subsection{\bf SSD Vees}

SSD Vees are reconstructed from pairs of oppositely charged
linked SSD-MPWC tracks originating from a common vertex. As the 
decays in this category occur close to the primary vertex, they 
tend to be lower momentum and accordingly have an excellent mass
resolution. These decays 
are principally $\kos$ and $\Lambda^0$ which decay upstream of the 
second SSD station. The secondary Vee vertex is required to 
lie downstream of the reconstructed SSD primary vertex. Basically,
these Vees
are found in the same way~\cite{lifet} that we find charm particles.
As such if  matching hits are found in the target silicon planes
then they are added into the silicon track definition.
The SSD Vees are the cleanest and 
the best defined category of Vees: the Vee track for this category
has a resolution comparable to that of two combined SSD tracks.
The standard deviations for the $\kos$ and $\Lambda^0$ mass 
distributions in this 
category are 3.6~MeV/$c^2$ and 1.6~MeV/$c^2$ respectively.

\subsection{\bf M1 Vees}

The M1 Vees are composed of $\kos$ and $\Lambda^0$ which 
decay between the last SSD plane and the first MPWC station, P0.
They are reconstructed with pairs of unlinked MPWC tracks and are
divided in three subcategories, according to the nature of their
components: {\it track-track, track-stub,} and {\it stub-stub}.

The reconstruction algorithm is substantially the same for the 
three topologies. For each candidate pair of unlinked MPWC tracks,
the intersection in the {\it xz} plane (non-bend) is first found; an
iterative procedure then traces the two prongs through the 
M1 field and determines the {\it y} location of the Vee vertex. In the
case of a track-stub Vee, the tracing also allows the computation
of the unknown momentum and charge of the stub prong. For a stub-stub 
Vee, it is necessary to further constrain the Vee vector to
originate from the SSD primary vertex and then the unknown
momenta can be computed. Finally, a global fit using the full
covariance matrices of the tracks (including multiple Coulomb
scattering effects) is applied to each Vee candidate to provide a
better estimate of the Vee decay vertex and the Vee momentum. 

M1 Vees are the most copious Vee category, accounting for over
70\% of the total reconstructed Vee sample.  
Their mass resolution and vertex resolution
are not as good as for the SSD Vees and are a very strong function
of the angle that the M1 field makes with respect to the normal
of the Vee decay plane. The resultant mass distributions often
have very non-Gaussian tails. To correct for this effect,  
Vees are retained based on a normalized mass cut or the 
difference between the reconstructed and nominal Vee mass divided 
by the anticipated resolution for a given topology. The normalized 
mass distribution is much closer to a true Gaussian distribution. 
Our selection criteria for the M1 Vees is that the absolute value 
of the normalized mass be less than 5. The $\kos$ mass resolution 
for the M1 Vee category varies from 6.6 MeV/$c^2$ for {\it stub-stub}
Vees to 5.9 MeV/$c^2$ for {\it track-track} Vees. 

\subsection{\bf RECON Vees}  

RECON Vees are Vees which decayed between P0 and P2. Because 
their decay region is further downstream, they tend to be
a Vee category with higher momentum.

RECON Vees are reconstructed using hits in P1, P2, P3, and P4
which have not already been used by the general MPWC pattern 
recognition (which only finds tracks with hits in P0).
This requirement significantly speeds up the algorithm, but
it reduces the efficiency in higher multiplicity events.
First, track projections in the {\it xz} (non-bend) view are constructed
and checked two at a time for intersecting between P0 and P2.
Projections which do not intersect in the desired region with any
other projection view are disregarded. Then, projections in
the U-view, V-view, and Y-view are formed and 
matched to the {\it xz} plane projections to
form tracks in space, with only loose requirements on the 
$\chi^2$/DOF of the track. Several track topologies are allowed:
P1234 (\emph{i.e.} tracks with hits in P1, P2, P3, and P4), P123, P234, and
P23. 
Finally, tracks are combined pairwise and a global 
fit to the Vee hypothesis is performed. The parameters of the 
fit are the five parameters for each track ({\it x} and {\it y} 
slopes and
intercepts and the {\it y} bend angle in M2) plus the coordinates
of the Vee vertex. In the case of P23 candidates (for which 
prongs are defined by a single point on each side of the magnet)
it is also necessary to assume that the Vee originates from the
SSD primary vertex in the target. RECON Vees sharing the same
{\it xz} projections are arbitrated on the basis 
of their $\chi^2$/DOF. 

This type of Vee is not used in FOCUS analyses.

\subsection{\bf P34 Vees}

P34 Vees are Vees which decayed between P2 and P3 in the magnetic
field of M2. Because their decay region is the most downstream
considered, they are the Vee category with the highest momentum.

P34 Vees are reconstructed using hits in P3 and P4 which have
not already been used both by the general MPWC pattern
reconstruction and by the hits used in the RECON Vee category.
Because the MPWC hits can be erroneously pre-assigned to another
track category, the reconstruction efficiency for P34 Vees is not high.
Track projections are formed in the {\it xz} (non-bend) plane and
checked two at a time for intersecting between P2 and P3. 
Projections which don't intersect in this region are discarded.
The projections from the U-view, V-view, and Y-view 
are formed and matched
to the X projections to form tracks. In order to determine 
the unknown momenta of the two station tracks, it is necessary
to further constrain the Vee vector to originate from the SSD
primary vertex. 

This type of Vee is not used in FOCUS analyses.

\subsection{\bf One-link SSD Vees}

One-link SSD Vees are reconstructed from the combination
of a linked SSD and a MPWC track with an unlinked SSD track. These 
are $\kos$ and $\Lambda^0$ which decayed before the second
SSD station, but for which one of the decay prongs falls out of the 
M1 geometrical acceptance and therefore is not found in the
PWC. 

The linked SSD to MPWC track and the unlinked SSD track are required
to make a good space vertex with a confidence level greater than 1\%.
A primary vertex is used as a constraint. The significance of 
separation between the primary 
and Vee decay vertex ($L/\sigma_L$,  is computed and it is required to have
$L/\sigma_L >~$10. Also the primary-secondary vector is required
to lie in the plane of the two decay prongs. By knowing the
momentum of the linked track, the primary-secondary direction
and the unlinked track direction, it is possible to balance
the transverse momentum  and compute the total momentum of the
unlinked track (there is no two-fold ambiguity in the kinematics).
Finally, the invariant mass of the two tracks is computed for the
$\kos$ hypothesis and for  the $\Lambda^0$ hypothesis. In the
$\kos$ case no \v{C}erenkov requirements in the two prongs are
imposed,
while in the $\Lambda^0$ case the linked prong is identified by
the \v{C}erenkov algorithm~\cite{wiss} to have a light pattern
more consistent with a proton hypothesis than a pion hypothesis.
The $\Lambda^0$ candidates
where the pion is linked and the proton is unlinked are not 
reconstructed, since this kind of Vee contained very little
signal over an overwhelming background. These Vees have
a $\kos$ mass resolution of 4.7 MeV/$c^2$.

\subsection{\bf MIC Vees}

MIC Vees have the decay vertex between the second and the 
fourth (last) SSD station. The reconstruction algorithm
starts by projecting unlinked MPWC tracks backward  onto the
SSD detector and by searching for unused hits in the last two stations.
If one or more matching triplets of hits are found, the parameters
of the track are recomputed by a global fit which uses both
the SSD and the MPWC hits. The new reconstructed tracks are then
checked two at a time to see if they originate from the same vertex in
space, and a cut on the distance of closest approach (DCA) is 
imposed. Candidate Vees sharing one prong are arbitrated on the 
basis of minimum DCA. The $\kos$ mass resolution for this category
is 4.4 MeV/$c^2$.  

This type of Vee is not used in FOCUS analyses.

\subsection{\bf Single-linked MPWC Vees}

Single-linked MPWC Vees share the same decay region of the
M1 Vees (between the SSD detector and P0), but are composed of one
linked SSD to MPWC track and one unlinked MPWC track. The
reconstruction algorithm is essentially the same as for the M1
Vees, and they are divided into three subcategories:
{\it track-track, track-stub,} and {\it stub-stub}. For the
$\kos$ reconstruction there was too much background in these
categories and they were never used. However, for the 
$\Lambda^0$ they were very important particularly when the
the $\Lambda^0$ was a decay product of a $\Xi^-$ or a $\Omega^-$.

\subsection{\bf Arbitration of Vee types}

In high multiplicity events it is possible for a track to 
be found in more than one Vee category.  This occurs when
one or both of the daughter tracks in a Vee are found in
other Vee candidates. The Vee candidates are arbitrated so 
that a single MPWC track is used just once in forming a Vee.
In order for a Vee to be arbitrated it must first have passed 
the basic requirements such as being within the mass windows
and satisfying the normalized mass cuts
and it must have a track shared with another Vee that satisfies
the same basic conditions. The decision to keep a Vee is made with the
following criteria.

$\bullet$ Two SSD Vees sharing a microstrip segment are arbitrated 
based on which
Vee has the smaller distance of closest approach at the decay
vertex location.

$\bullet$ Two M1 {\it track-track} Vees sharing a MPWC track
  are arbitrated based on which
Vee has the smaller distance of closest approach at the decay
vertex location. 

$\bullet$ M1 {\it track-track} Vees are always selected 
over M1 {\it track-stub} Vees.

$\bullet$ Two M1 {\it track-stub} Vees sharing a MPWC track
are arbitrated based on which
Vee has the smaller $\chi^2$/DOF of the fit.

$\bullet$ M1 {\it track-stub} Vees are always selected over 
M1 {\it stub-stub} Vees.

$\bullet$ Two M1 {\it stub-stub} Vees sharing a MPWC track
are arbitrated based on which
Vee has the smaller $\chi^2$/DOF of the fit.

\subsection {\bf Vee Refit}

Vees are reconstructed with only a rough knowledge of their
actual production vertex. For the purposes of track finding 
the assumption is made that the tracks originate at the 
primary vertex (see Fig. \ref{fig:refitdecay}). When 
a secondary charmed particle vertex is identified, then the Vee is 
refit to the true production vertex. This refit slightly 
improves the Vee production angle and the Vee mass.

\begin{figure}[htbp]
\begin{center}
{\includegraphics[width=7.5cm]{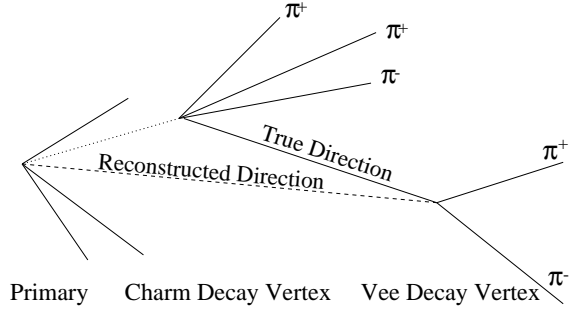}}
\end{center}
\caption{A sketch of the charmed particle decay 
$D^+\rightarrow K^0_S\pi^+\pi^+\pi^-$ where the 
$K^0_S\rightarrow \pi^+\pi^-$
is presented to display the need for a Vee refit.
The $K^0_S$ is first reconstructed to originate
from the primary vertex, but a refit is necessary when it is
reassigned to come from a new vertex.}
\label{fig:refitdecay}
\end{figure}

Six histograms of $K^0_S$ Vee types are presented in 
Fig. \ref{fig:ksmulti}. The histograms are all plotted on the same
mass scale and were made from the same small data sample (approximately
0.5\% of the total). The
({\emph{M1 stub-stub}}) Vees have the most background and the 
poorest mass resolution. The ({\emph{M1 track-track}}) and 
({\emph{M1 track-stub}}) Vees are the most common categories. 
The ({\emph{MIC}}) Vees have the least background, but the yield
is comparatively low. The 
({\emph{SSD}}) Vees have the best mass resolution, but can have
considerable background since the decay occurs close to the
interaction region. The ({\emph{One-link}})
Vees also have good mass resolution due to their very low momentum, 
but these Vees have considerable
background.

\begin{figure*}[htbp]
\begin{center}
{\includegraphics[width=4.5cm]{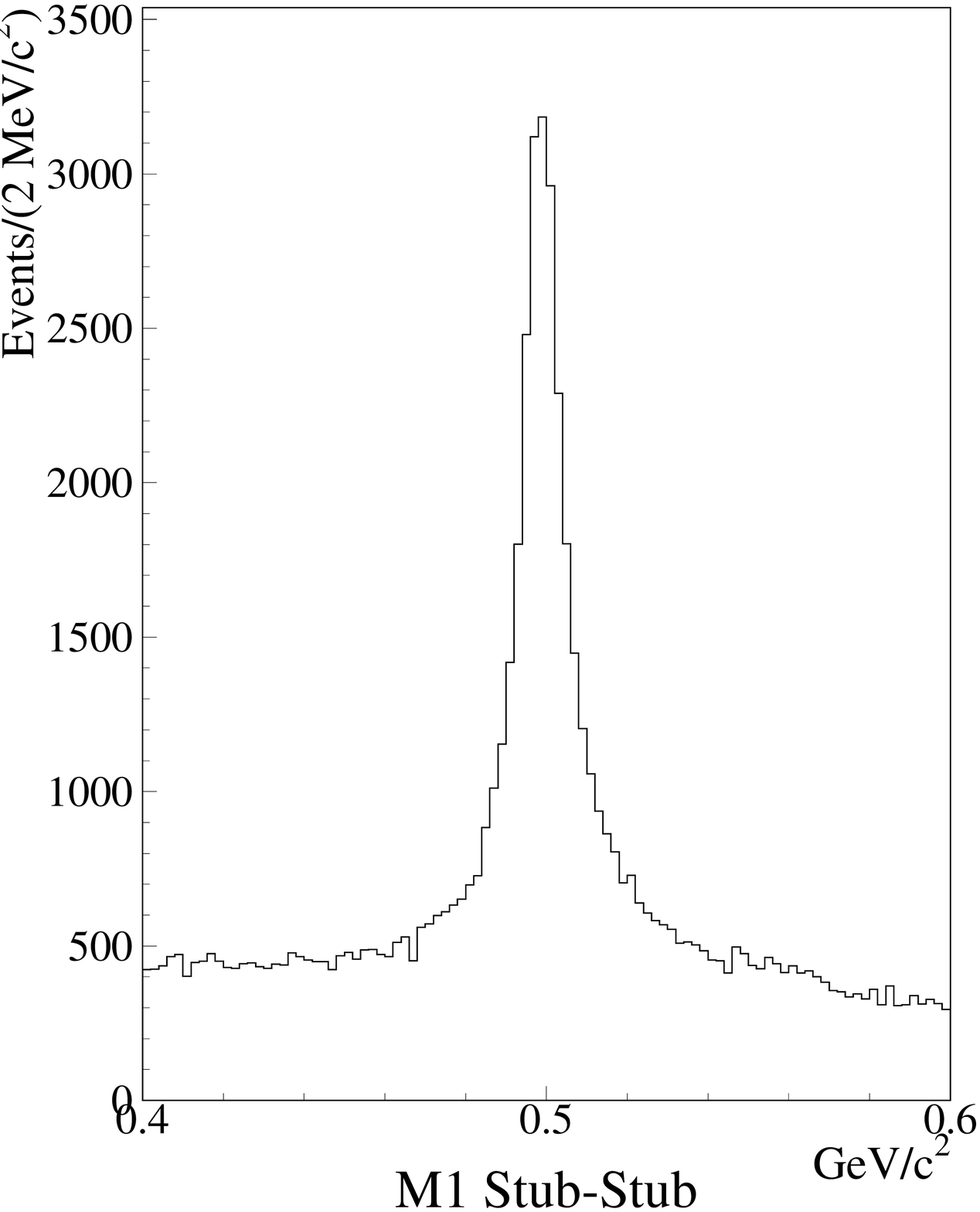}}
{\includegraphics[width=4.5cm]{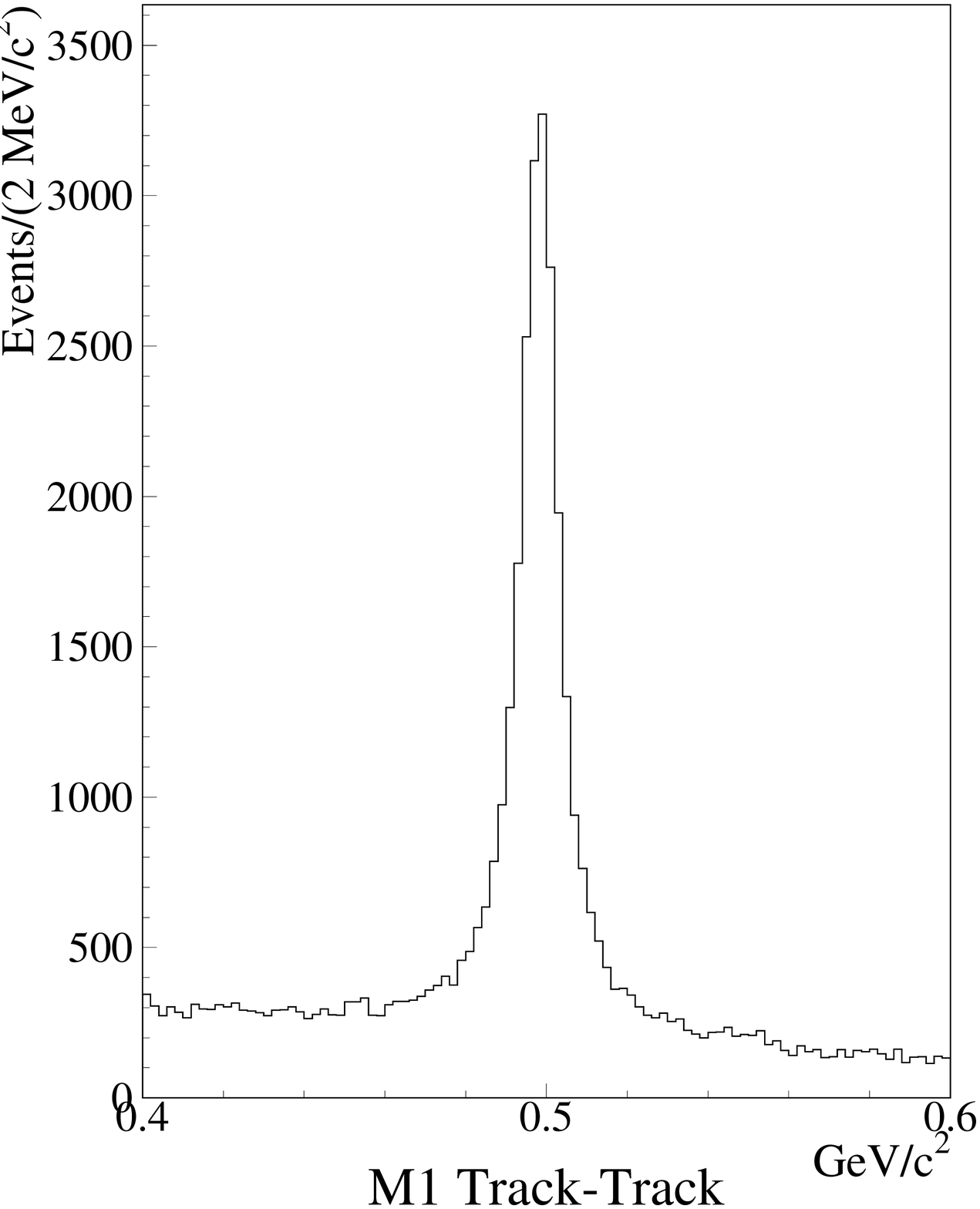}}
{\includegraphics[width=4.5cm]{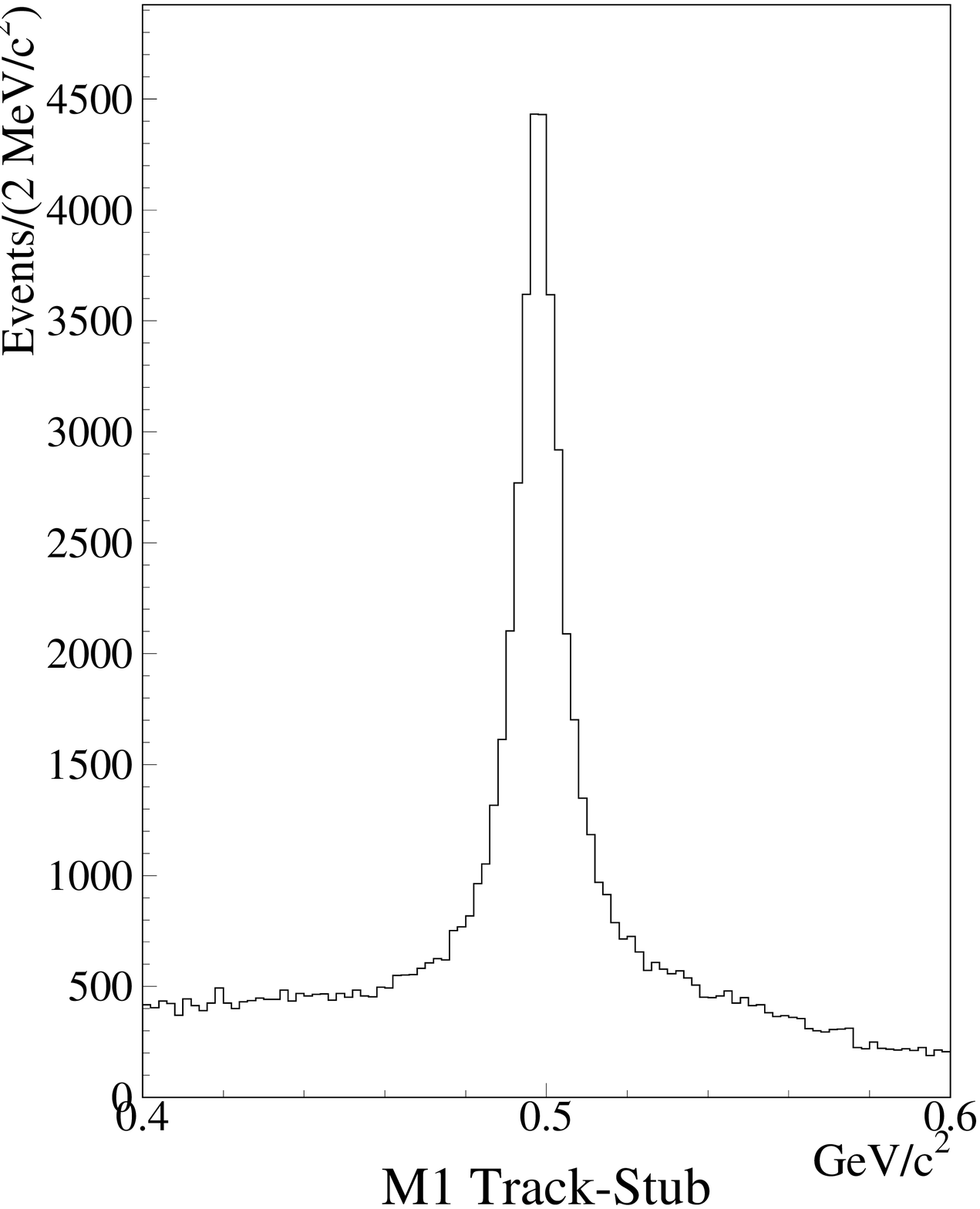}}
{\includegraphics[width=4.5cm]{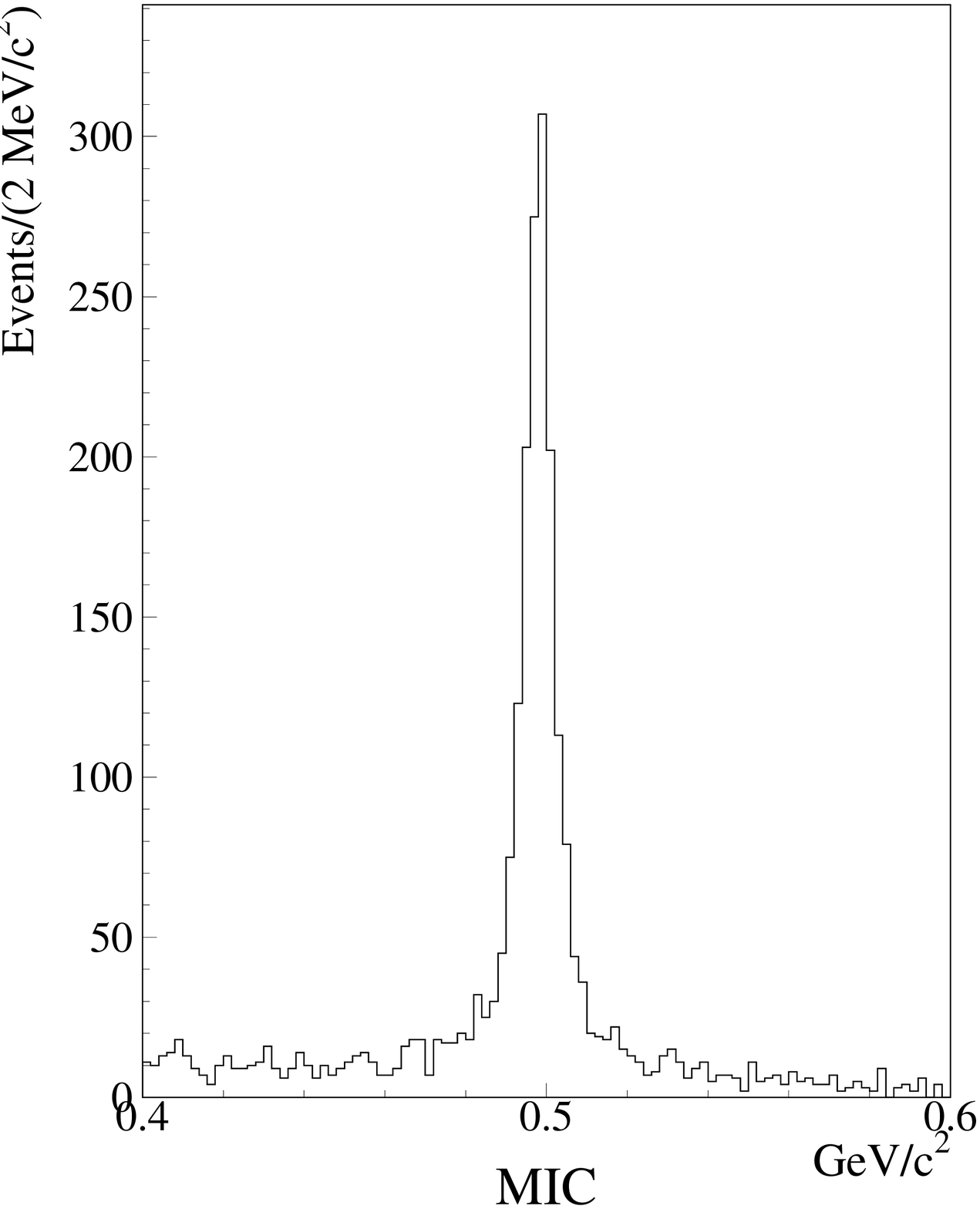}}
{\includegraphics[width=4.5cm]{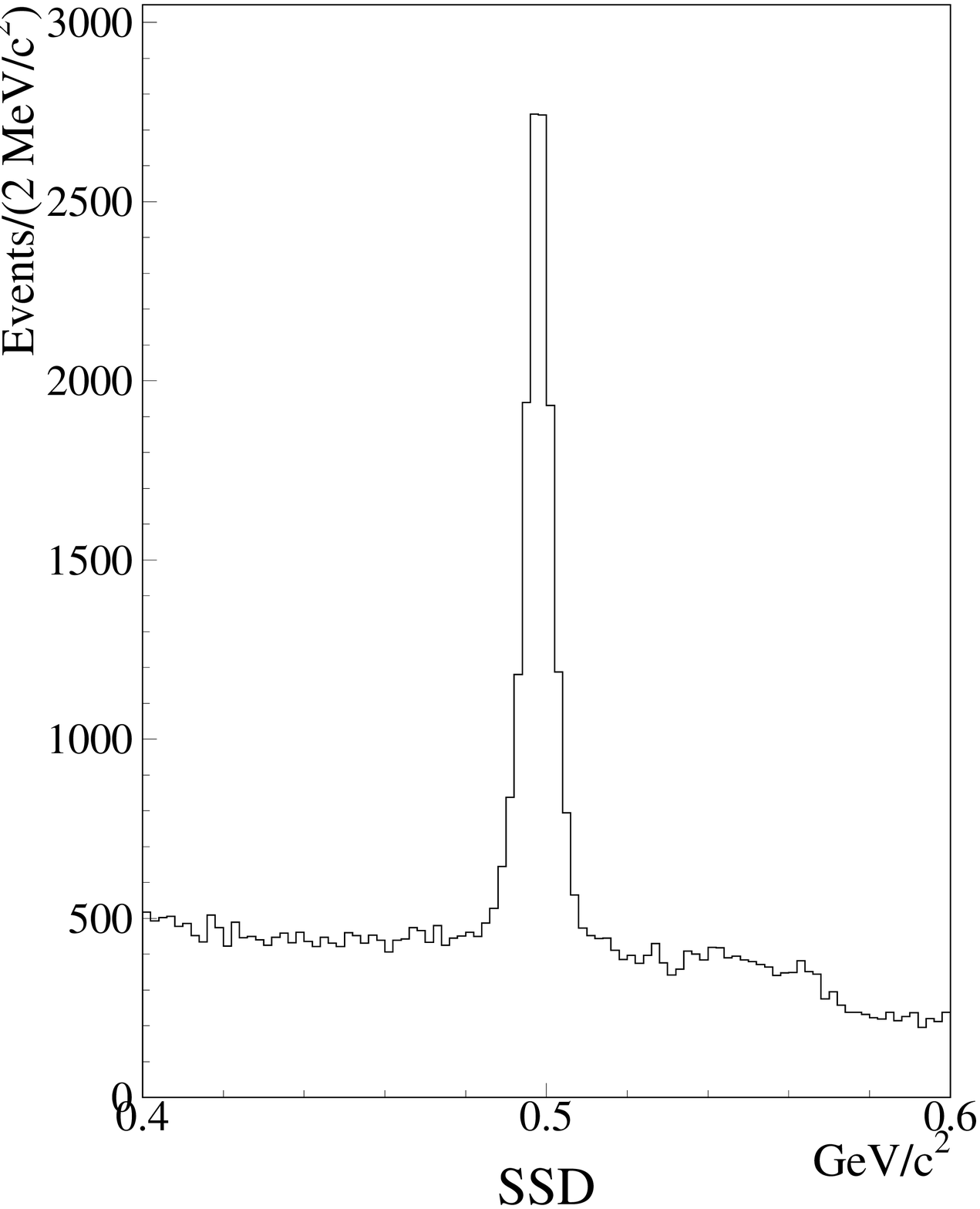}}
{\includegraphics[width=4.5cm]{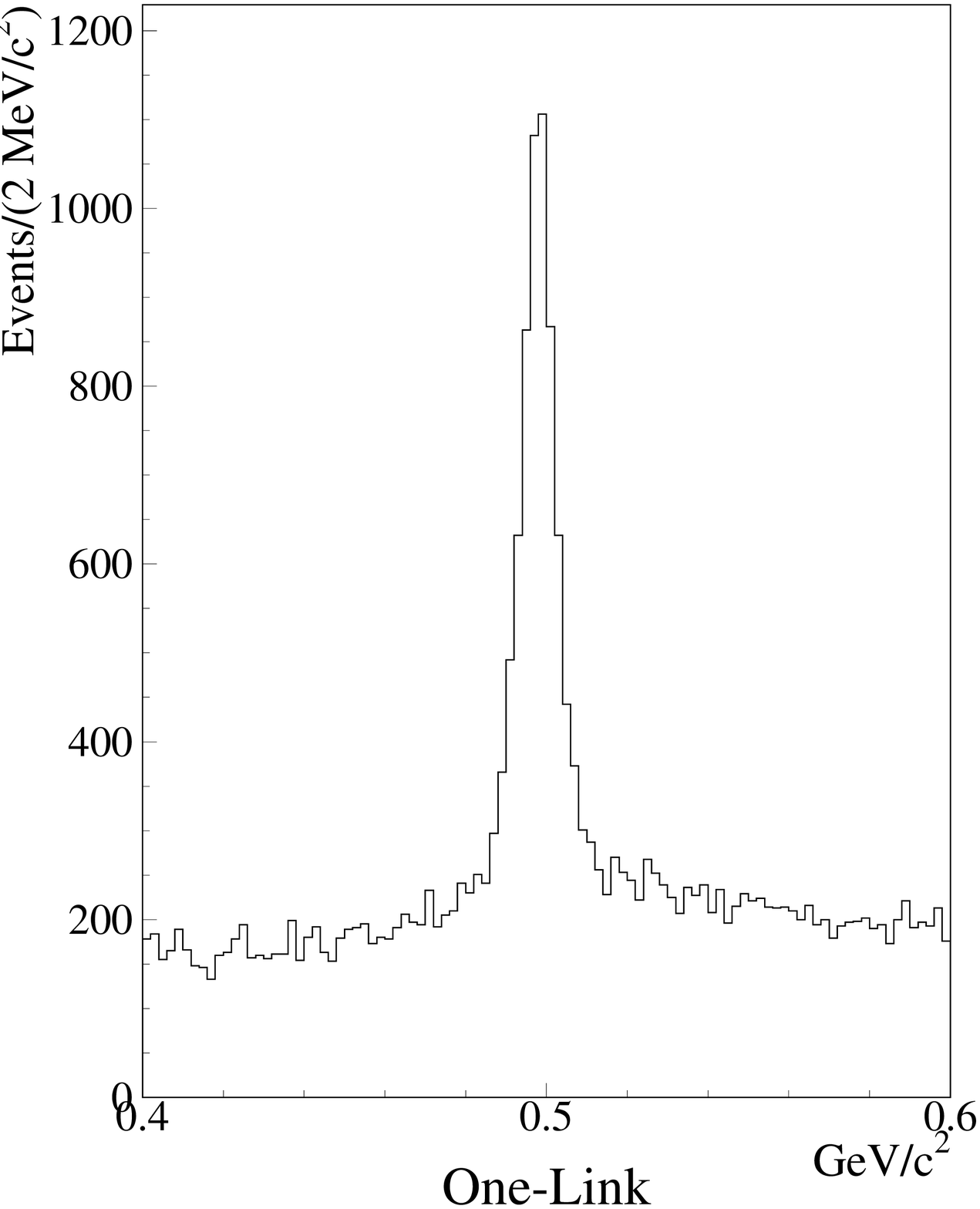}}
\end{center}
\caption{Histograms of a sample of six categories
of $K^0_S$'s which
are used in the analysis. The dominant categories in charm
decays are the 
({\it M1 track-track}) and ({\it M1 track-stub}) Vees. The
categories with the best resolution are {\it MIC}, {\it SSD},
and {\it One-link} Vees.}
\label{fig:ksmulti}
\end{figure*}

Nine histograms of 
$\Lambda^0$ Vee types are plotted in Fig. \ref{fig:lbmulti}.
The data represents about 0.5\% of the total $\Lambda^0$ yield. 
For the $\Lambda^0$ decays it is useful to 
retain the ({\emph{Single-linked}}) categories,
since $\Xi^-$'s and $\Omega^-$'s decay to $\Lambda^0$'s and occasionally
the proton in the $\Lambda^0$ decay is erroneously linked with 
the $\Xi^-$ or $\Omega^-$ track segment in the SSD. 

\begin{figure*}[htbp]
\begin{center}
{\includegraphics[width=4.5cm]{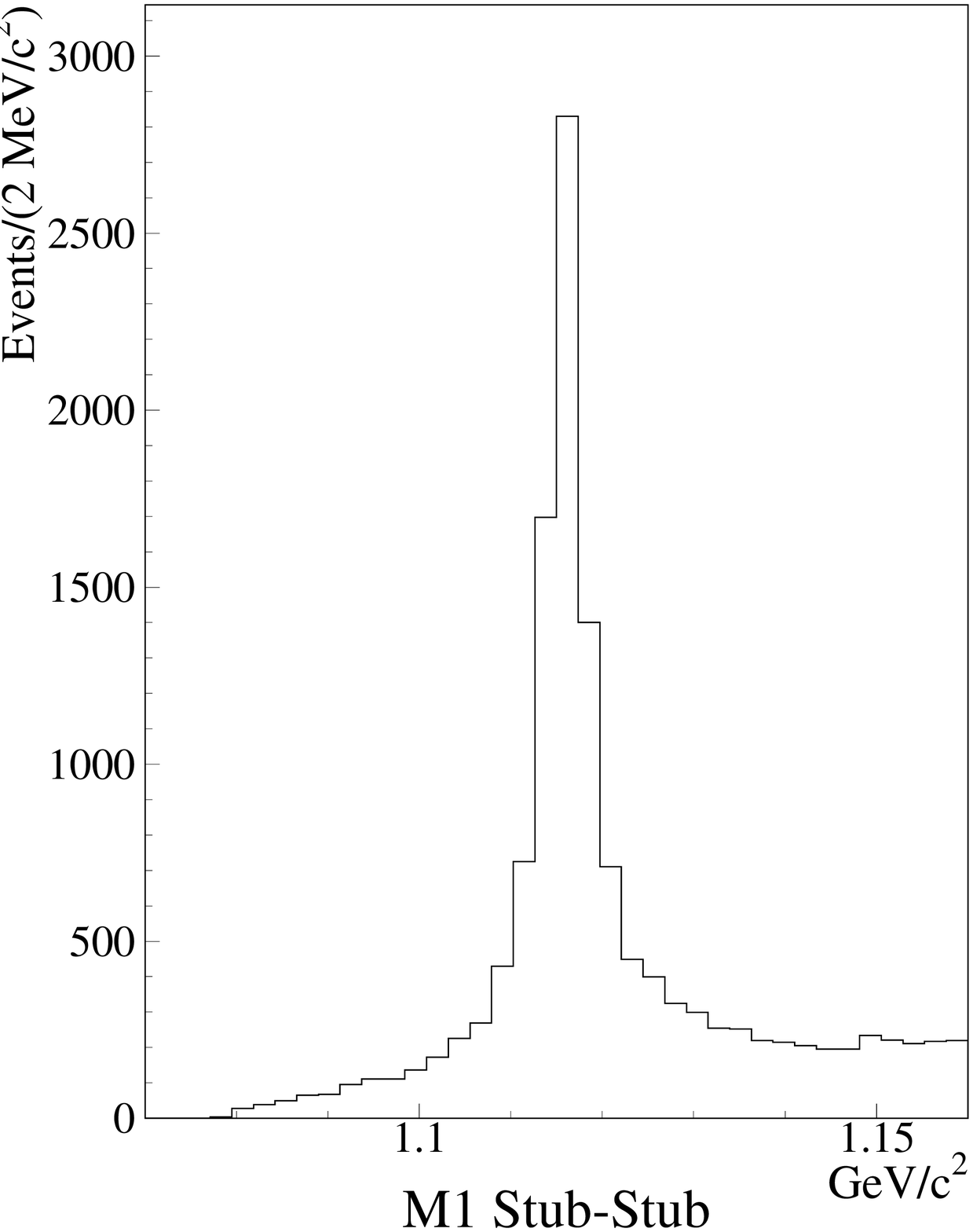}}
{\includegraphics[width=4.5cm]{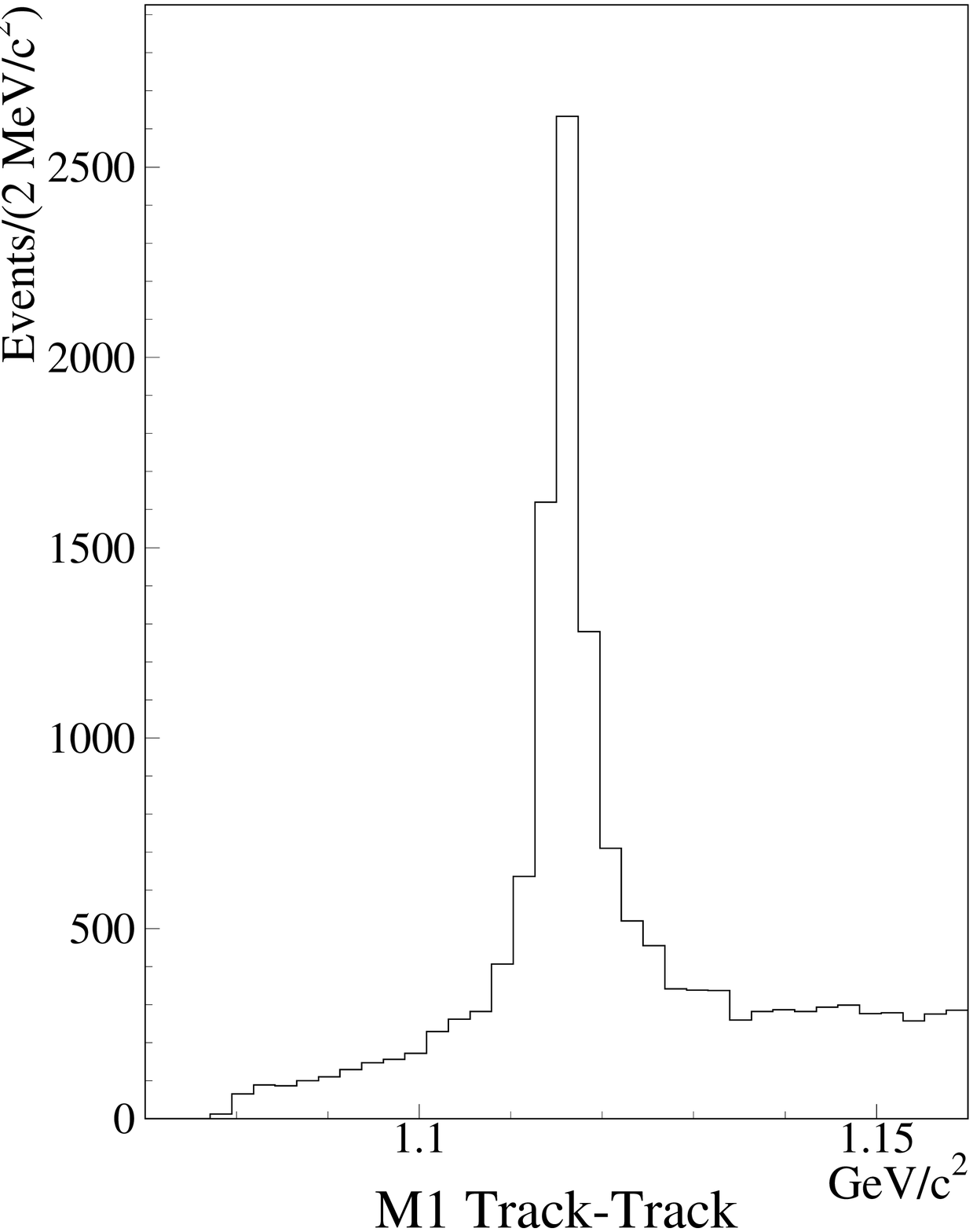}}
{\includegraphics[width=4.5cm]{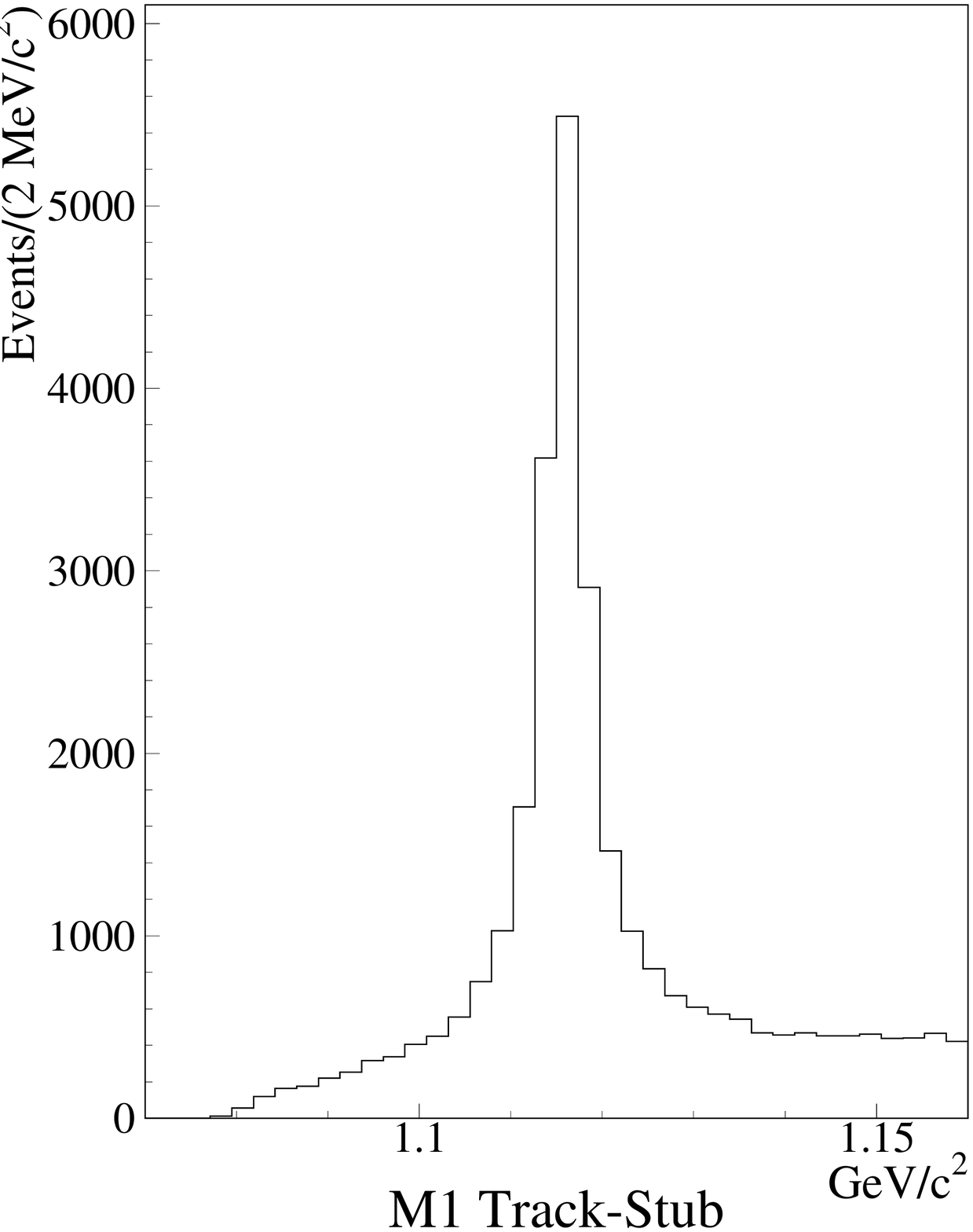}}
{\includegraphics[width=4.5cm]{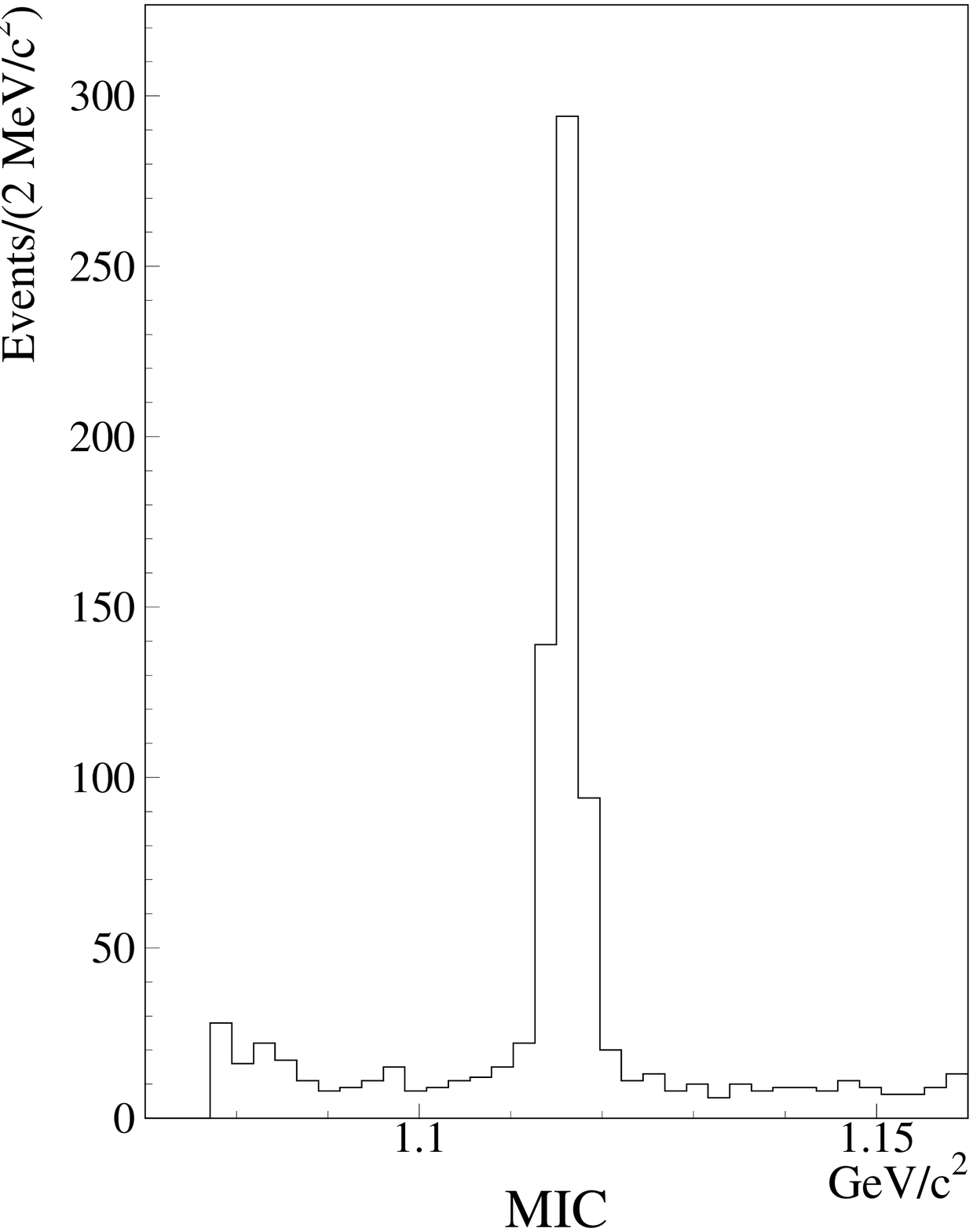}}
{\includegraphics[width=4.5cm]{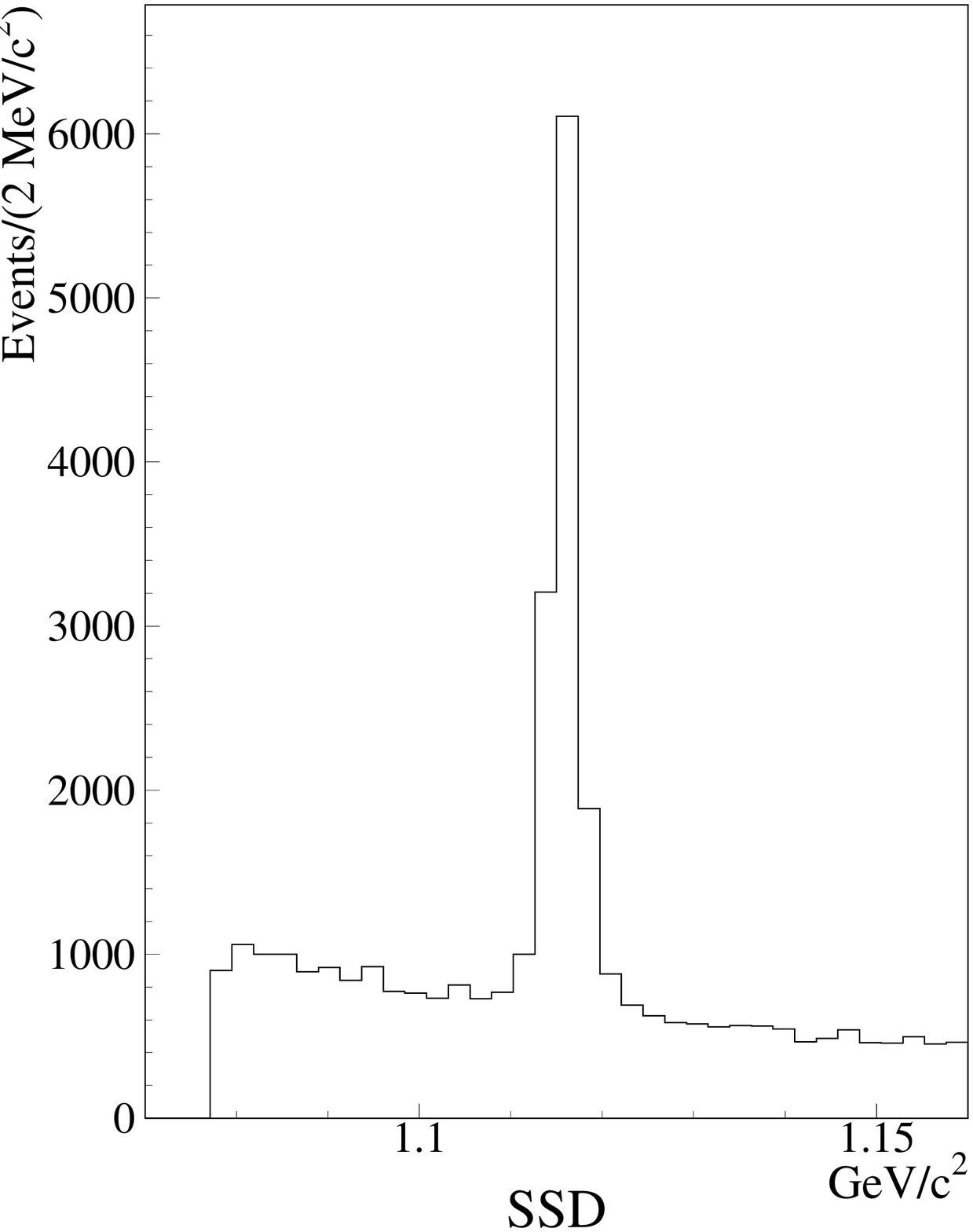}}
{\includegraphics[width=4.5cm]{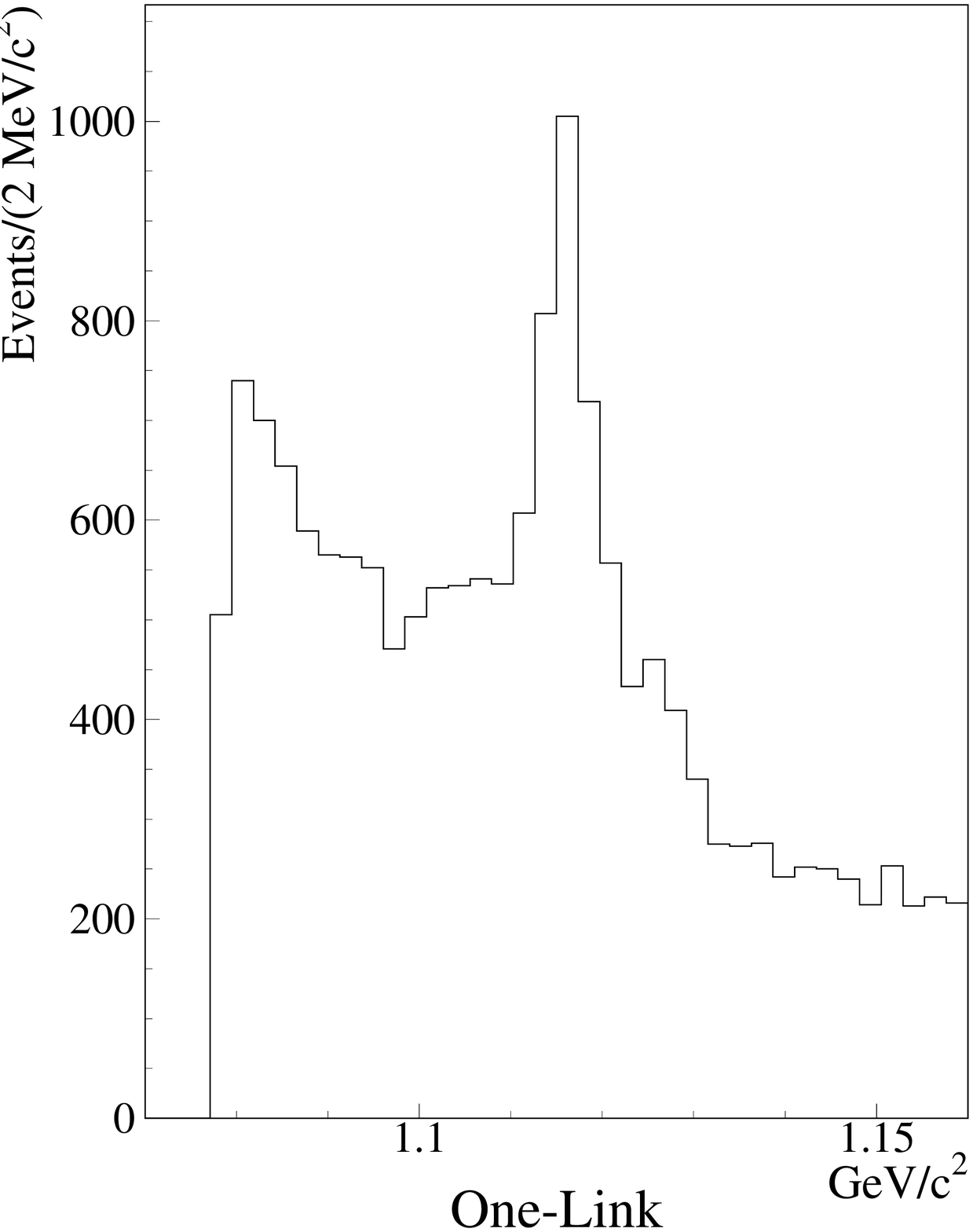}}
{\includegraphics[width=4.5cm]{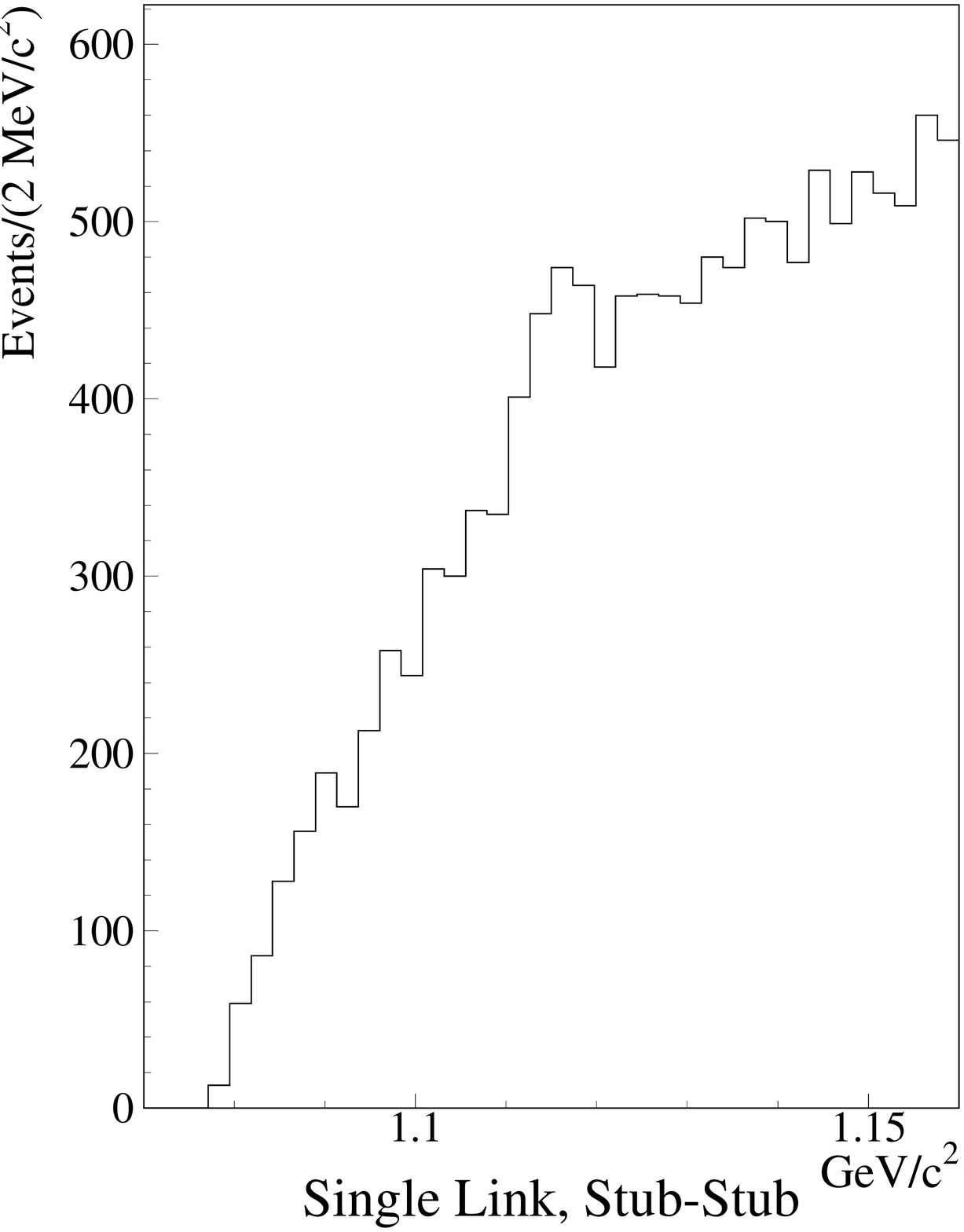}}
{\includegraphics[width=4.5cm]{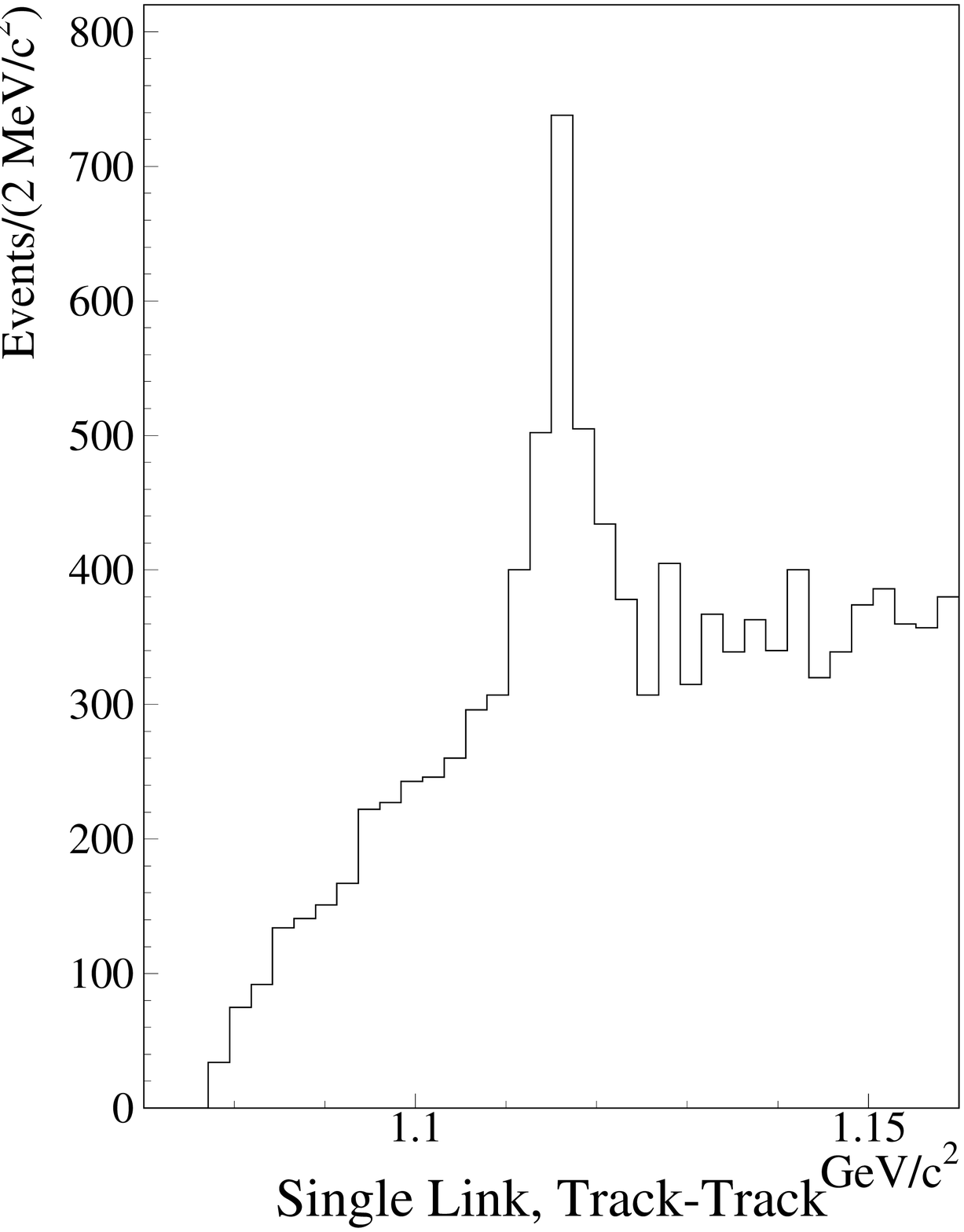}}
{\includegraphics[width=4.5cm]{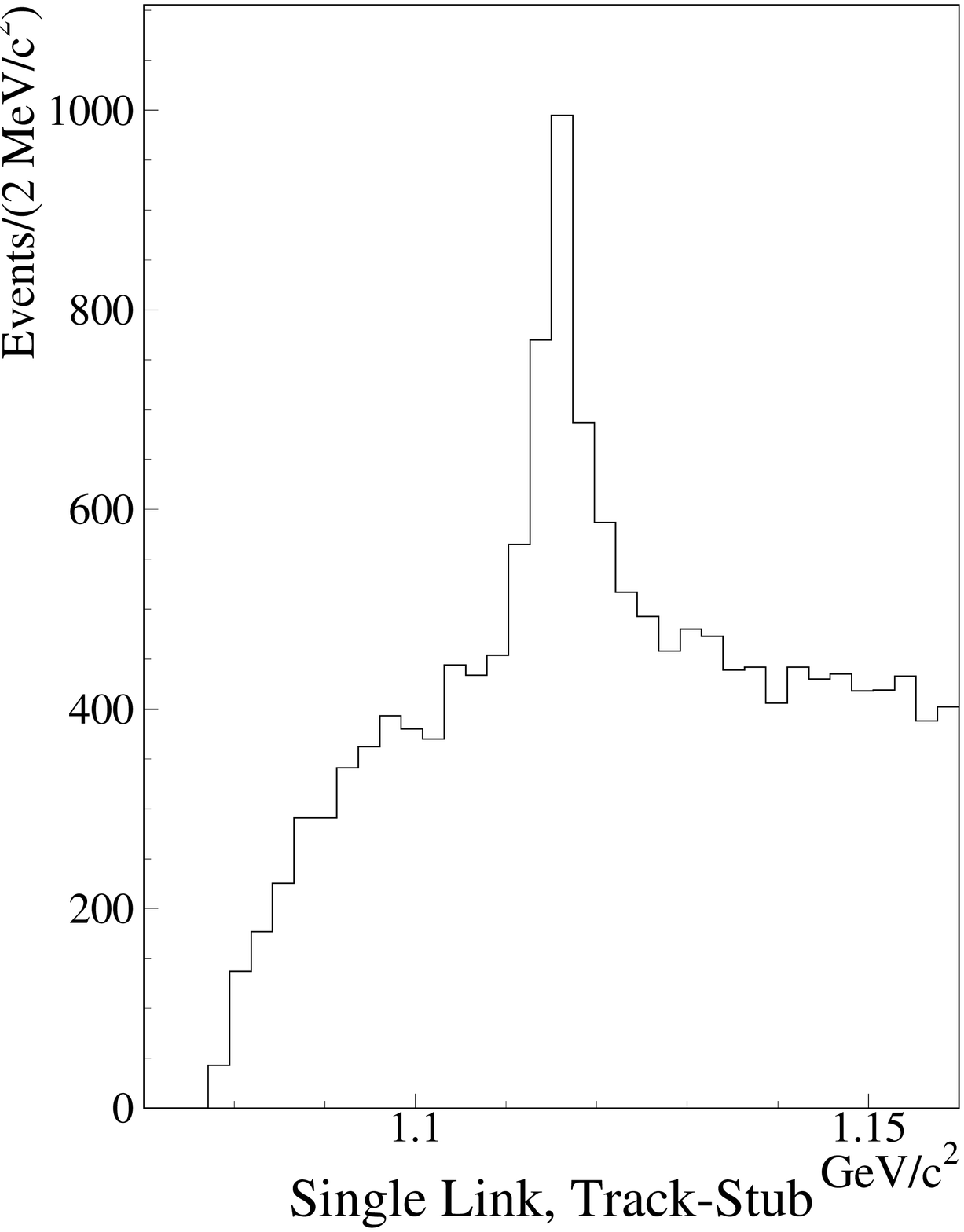}}
\end{center}
\caption{Histograms of a sample of nine categories
of $\Lambda^0$'s which
are used in the FOCUS analyses. The final row of Single-linked
$\Lambda^0$'s are not used in charm analyses involving 
direct charm decays to $\Lambda^0$'s, but account for almost 
15\% of the $\Xi^-$'s and $\Omega^-$'s decays.}
\label{fig:lbmulti}
\end{figure*}

\section{\bf Kinks}
The term ``Kink" refers to the topology where one charged particle
decays to another charged particle and a neutral particle. The
neutral particle
is undetected in the tracking detectors.

A schematic of a Kink decay is given in Fig.~\ref{fig:kinkscem}. The parent 
particle with measured direction cosines, $\alpha, \beta, \gamma$
and with mass, $m_p$, decays to particles $m_1$ and $m_2$. Particle
$m_1$ has its momentum and direction cosines measured. Particle 2
is neutral and goes undetected in the spectrometer. Through kinematic
constraints and by assuming its mass the parent momentum is calculated
to within a twofold ambiguity.

\begin{figure}[htbp]
\begin{center}
{\includegraphics[width=13cm]{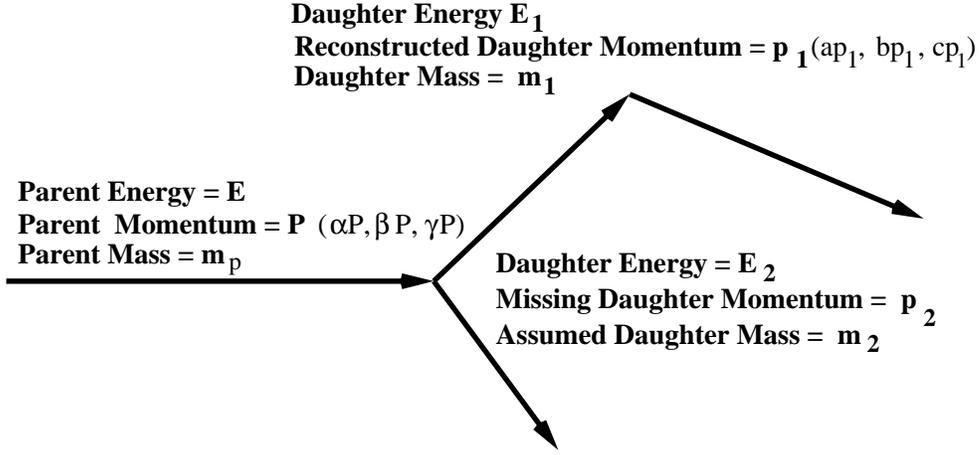}}
\end{center}
\caption{A sketch of the Kink decay process where a parent
particle decays to two tracks where one track is completely 
measured and the other track information is unknown. The direction
of the parent track is known, but its momentum is unknown.
Through kinematics the momentum of the parent particle can be
calculuated to within a twofold ambiguity.}
\label{fig:kinkscem}
\end{figure}

The momentum of $p_2$ is determined through conservation of momentum,

\begin{equation}
{\bf p_2} = (\alpha P - ap_1, \beta P - bp_1, \gamma P - cp_1) 
\end{equation}
and the square of the energy  of $E_2$ is given by

\begin{equation}
E_2^2 = (E - E_1)^2 = p_2^2 +m_2^2
\end{equation}

Substituting the expression for ${\bf p^2_2}$ we obtain 

\begin{equation}
(m_p^2 + m_1^2 - m_2^2) + 2Pp_1(\alpha a + \beta b + \gamma c) = 2 EE_1
\end{equation}

Squaring this equation, substituting $P^2 = E^2 - m^2_p$,  and 
solving for $P$, we find

\begin{equation}
P = {{MC \pm \sqrt{M^2C^2 - (1-C^2)(m^2_p - M^2)}} \over {(1-C^2)}}
\end{equation}

where $M = {{m^2_p + m^2_1 -m^2_2} \over {2E_1}}$, 
$C = {{p_1(a\alpha + b\beta + c\gamma)} \over {E_1}}$, and $P > 0$. 
Thus, we find the twofold ambiguity. 
 
The three decays $\Sigma^+\rightarrow p\pi^0$,
$\Sigma^+\rightarrow n\pi^+$, and $\Sigma^-\rightarrow n\pi^-$
are reconstructed using the {\it Kink} algorithm. It should
be noted that the $\Sigma^+$($suu$) and the $\Sigma^-$($sdd$)
are {\it not} charge conjugate partners and that the 8~MeV/$c^2$
difference in their masses is important to include in the algorithm.
The $\Sigma^+$
decays weakly to $p\pi^0$ 51.6\% of the time and to $n\pi^+$
48.4\% of the time. The $\Sigma^-$ decays to $n\pi^-$
essentially 100\% of the time. A complete list of the
Kink candidates considered is given in Table \ref{tab:kinktab}.
The fraction of the meson decays that we would recover from the
Kink algorithm is insignificant compared to the prevalence of
the non-decaying ones. We therefore concentrated on the hyperon
decays, $\Sigma$'s, $\Xi$'s, and $\Omega$'s.
Initially, we concentrated on the $\Sigma$ decays.
The decays $\Xi^-\rightarrow \Lambda\pi^-$, $\Omega^-\rightarrow \Lambda
K^-$, and $\Omega^-\rightarrow \Xi^0\pi^-$
will be further considered in Section 6.

\begin{table}[htbp]
\caption{Decay topologies considered in the Kink algorithm.
\label{tab:kinktab}}
\vspace{0.4cm}
\begin{center}
\begin{tabular}{clcl}
\hline
Type& Decay & c$\tau$(cm) & Comments \\
\hline
 1 & $K^-\rightarrow \mu^- \nu$ & 371.3 &muon identified \\
 2 & $\Sigma^+\rightarrow n~\pi^+$ & 2.396 & neutron energy found \\
   &                               & &in hadron calorimeter\\
 3 & $K^-\rightarrow \pi^- \pi^0$ &  371.3 & \\
 4 & $\Sigma^+\rightarrow p~\pi^0$ & 2.396 & $p$ identified by
\v{C}erenkov \\
 5 & $\Sigma^-\rightarrow n~\pi^- $ & 4.434 &neutron energy found \\
   &                               & & in hadron calorimeter\\
 6 & $\pi^-\rightarrow \mu^- \nu$ & 780.45 &muon identified \\
 7 & $\Xi^-\rightarrow \Lambda\pi^-$ & 4.91 &                   \\
 8 & $\Omega^-\rightarrow \Lambda K^-$ & 2.46 & $K$ identified by
\v{C}erenkov \\
 9 & $\Omega^-\rightarrow \Xi^0\pi^-$ & 2.46 & \\
\hline
\end{tabular}
\end{center}
\end{table}

The Kink algorithm
begins by looping over each unlinked microstrip track that points
into the M1 aperture and by pairing it with every unlinked
MWPC track which also points into the M1 aperture. MWPC tracks
used in Vee candidates 
 are not considered. A preliminary Kink
vertex location is determined by intersecting the microstrip
and MWPC tracks in the {\it xz} plane. This choice provides a rough
estimate of the {\it x} and {\it z} location, which is required to be
downstream of the last microstrip station and upstream of P0.

If the MWPC track passes through all five MPWCs,
then its momentum is already
determined from M2 and it can be traced through the field of
M1 to the estimated Kink {\it z} position. If the Kink
{\it z} coordinate
is upstream of M1, then  the {\it y} distance between the
projected 5-chamber track and the microstrip track is compared
and required to be less than 2.5~mm. This helps eliminate
spurious Kink candidates. The parent momentum can only be
calculated by making a particular decay hypothesis and by solving
the kinematic equations. This involves assuming the parent
($\Sigma^{\pm}$) and daughter masses (including the missing
neutral daughter) and balancing the momentum transverse to the
parent direction. This results in a two-fold ambiguity in the
parent momentum.

If the Kink {\it z} position is within M1, then the ambiguity can be
broken and a unique solution is found. The MWPC track is traced
upstream to the {\it z} of the Kink.
The {\it x} and {\it y} coordinates of the Kink are then given
by the traced
position of the track at this {\it z}. The microstrip track is then traced
downstream through M1 to the Kink position. The trace is iterated
several times, using trial momenta, until a momentum is found
which best traces the microstrip track to the Kink vertex.
The transverse momentum is balanced and the kinematic equations
are solved. If there are two physical solutions for the
parent momentum, then the one nearest to the momentum calculated
by the iterative trace is used. In theory, one could avoid solving
the Kink kinematic equations for decays in the magnetic field
by simply using the microstrip trace
momentum, however often
very little
magnetic field is traced through and the resolution on the microstrip
track curvature (momentum) is not well defined. For these reasons
the kinematic solutions were always used to determine the momentum.

For 3-chamber MWPC tracks we consider only the case where
the Kink position
is upstream of M1. The {\it x} and {\it y} positions of the
microstrip track at the previously estimated Kink {\it z} position
are used as the Kink vertex location. An iterative trace procedure,
similar to the one discussed above, is used to find the best
momentum for the 3-chamber track. Unfortunately, a two-fold
ambiguity exists in the parent momentum calculation.

All putative Kinks are reconstructed according to all three decay
hypotheses (types 2, 4, and 5) from Table \ref{tab:kinktab}.
Some hypotheses are rejected on the basis of
particle identification of the charged daughter or on the basis
of calorimetry information. Charged particle identification
is made by re-running the \v{C}erenkov code with the momentum
determined from the Kink reconstruction in order to obtain a
new \v{C}erenkov identification estimate ({\it i.e.}
the momentum changes). For the $\Sigma^+\rightarrow p\pi^0$
hypothesis the \v{C}erenkov light pattern is required to be
more consistent with a proton assumption than the pion
assumption by a factor of~7.
For the $\Sigma^+ \rightarrow n\pi^+$ and $\Sigma^- \rightarrow
n\pi^-$ hypotheses the pion is required to not be consistent
with being an electron, kaon, or proton and the
neutron must impact the hadron calorimeter. The neutron
must deposit sufficient
energy, E,  surrounding the point where the neutron is expected
to strike the calorimeters and must satisfy 0.3 $<$ E/p $<$ 2.0,
where p is the momentum of the neutron calculated from
the Kink kinematic equations. A detailed description of the 
FOCUS hadron calorimeter and its performance  is found in 
Reference~7.

There is no method to plot the mass of the reconstructed Kinks
directly because of the assumed mass constraint in the algorithm.
To show the success of our algorithm we present the invariant
mass plots in Fig. \ref{fig:riccardi} for $\Lambda^+_c\rightarrow
\Sigma^-\pi^+\pi^+$ where the $\Sigma^-$ is reconstructed through
the $n\pi^-$ channel and for $\Lambda^+_c\rightarrow\Sigma^+\pi^+\pi^-$
where the $\Sigma^+$ is reconstructed through the $p\pi^0$ channel
and through the $n\pi^+$ channel. Care has been taken to weight the
cases where there are two solutions such that the yields in 
the caption for Fig. \ref{fig:riccardi} are correct.

\begin{figure*}[htbp]
\begin{center}
{\includegraphics[width=14cm]{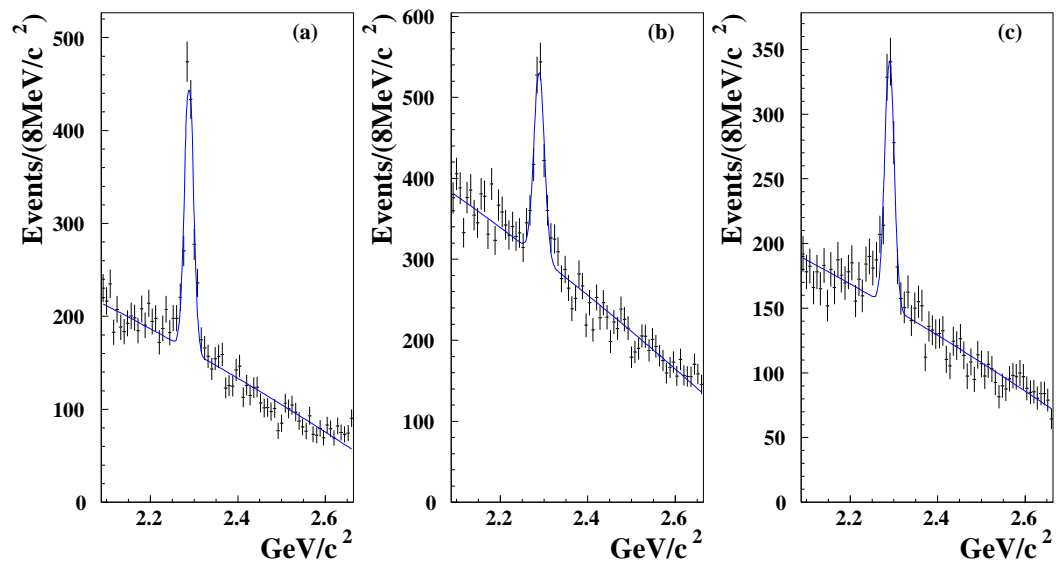}}
\end{center}
\caption{Three invariant  mass plots for the decays
(a) $\Lambda^+_c\rightarrow\Sigma^+\pi^+\pi^-$ where
$\Sigma^+\rightarrow p\pi^0$
with a yield of 915$\pm$50 events,
(b) $\Lambda^+_c\rightarrow\Sigma^+\pi^+\pi^-$ where
$\Sigma^+\rightarrow n\pi^+$
with a yield of 854$\pm$65 events,
and for (c) $\Lambda^+_c\rightarrow\Sigma^-\pi^+\pi^-$
where $\Sigma^-\rightarrow n\pi^-$ with a yield of 654$\pm$42 events.}
\label{fig:riccardi}
\end{figure*}

\section{\bf $\Xi^-$'s and $\Omega^-$'s}

The $\Xi^-\rightarrow \Lambda^0\pi^-$ and $\Omega^-\rightarrow
\Lambda^0K^-$ decays are reconstructed via several
techniques. 
The $\Xi^-$'s and $\Omega^-$'s decaying upstream of the silicon
microstrip detector are reconstructed differently than those
decaying downstream of the microstrip. If a Vee is not found,
or if the $\Lambda^0-$~track vertex is located downstream of the
Vee vertex ({\emph{i.e.}} due to poor vertex resolution in the
event), then we use a technique referred to as `multivees'
where there are three unlinked tracks. Finally, if no fully
reconstructed $\Xi^-$ or multivee is found, then we use a
Kink algorithm to find $\Xi^-\rightarrow \Lambda^0\pi^-$ where
the $\Lambda^0$ is unidentified. This algorithm is needed to 
reconstruct the $\Xi^-$ when $\Lambda^0\rightarrow n\pi^0$. 
Each of these techniques
will be described below. 

The fully reconstructed algorithm uses a common set of
requirements to select Vees as $\Lambda^0$'s. The Vee daughter
track with the highest momentum is considered the proton for the
$\Lambda^0$ hypothesis and the \v{C}erenkov algorithm is executed
using momenta of the Vee tracks as determined by the Vee 
algorithm. 

\subsection{\bf Upstream Reconstructed $\Xi^-$'s and $\Omega^-$'s}

{\it Upstream} $\Xi^-$'s and $\Omega^-$'s are those which decay
upstream of the first microstrip station ({\it i.e.} within the
target or between the target and the microstrip detector). They
are also referred to as `Type~1' decays. 
A schematic of a typical $\Xi^-$ decay in the category is presented
in Fig. \ref{fig:cas1decay}.

Linked
MWPC tracks are paired with each $\Lambda^0$ Vee which satisfies
the $\Lambda^0$ hypothesis.
Upstream decays are reconstructed by intersecting the $\Lambda^0$
vector and the MWPC track and by demanding that the confidence
level of this vertex be greater than 1\% and that the
$\Xi^-/\Omega^-$ be consistent with originating from a
production vertex further upstream. The distance between the
production vertex and the $\Xi^-/\Omega^-$ decay vertex is
defined as $L$. The track is assigned the pion mass to form the 
invariant mass for $\Xi^-$ hypothesis and is assigned the
kaon mass for the $\Omega^-$ hypothesis. There are two
requirements which are used to significantly improve the 
signal-to-noise for these $\Xi^-/\Omega^-$ signals. The first 
is the {\it significance of separation} of the $\Xi^-/\Omega^-$
decay vertex from its production vertex. and the second
is the isolation of the $\Xi^-/\Omega^-$ decay vertex from other 
tracks.

\begin{figure}[htbp]
\begin{center}
{\includegraphics[width=7.5cm]{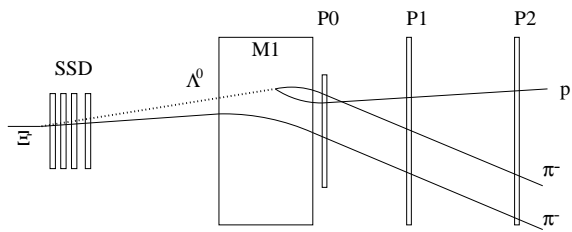}}
\end{center}
\caption{A schematic drawing in the bend view of the 
spectrometer of a $\Xi^-$ decay which
occurs upstream of the silicon strip detector (SSD).
Only the front part of the spectrometer is displayed.}
\label{fig:cas1decay}
\end{figure}

The {\it significance of separation} of the $\Xi^-/\Omega^-$ 
decay vertex from its production vertex is defined as the
quantity $L$/$\sigma_L$, where $\sigma_L$ is the error on 
$L$. By increasing the cut on $L$/$\sigma_L$ ({\it i.e.} $L$/$\sigma_L
>$1,~2,~3, etc.) one obtains a cleaner sample. The basic algorithm
requires $L$/$\sigma_L~>0$.

The isolation of the $\Xi^-/\Omega^-$ decay vertex from other 
tracks is tested by attempting to place other tracks in the 
vertex and by refitting it. A cleaner signal is obtained by
requiring that the confidence level from the fit to the
new vertex be less than a certain value. Invariant
mass plots for upstream decays for the  $\Lambda \pi^-$ and
$\Lambda K^-$ combinations 
are presented in Fig. \ref{fig:castype1}. Note that there is
considerable background under the $\Omega^-$ signal which is almost
entirely due to pion particle misidentification from the
$\Xi^- \rightarrow \Lambda \pi^-$ decays. One should also note that
$\Omega^-$ yield is about a factor of 20 less than the $\Xi^-$
yield.

\begin{figure}[htbp]
\begin{center}
{\includegraphics[width=6.5cm]{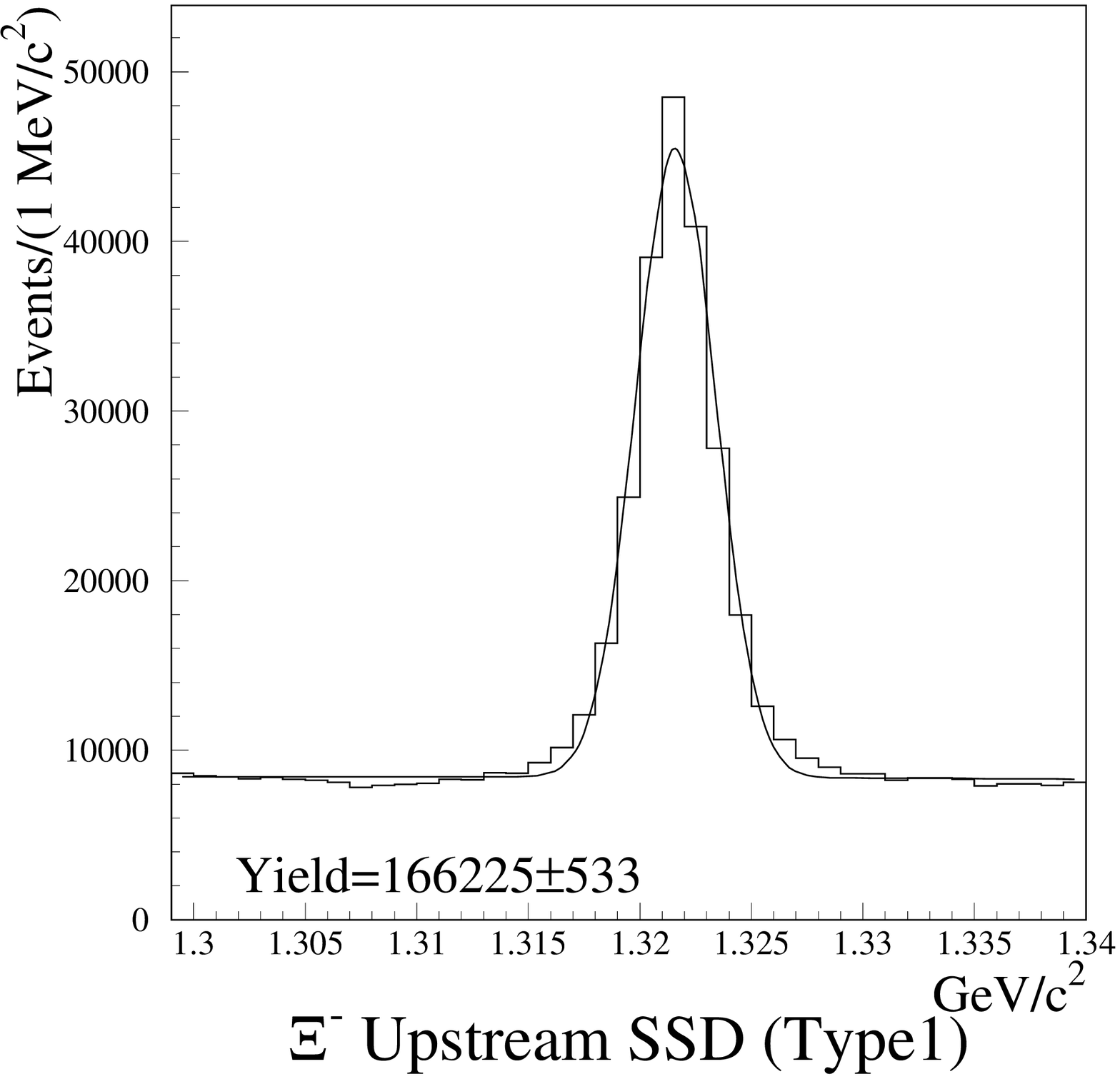}}
{\includegraphics[width=6.5cm]{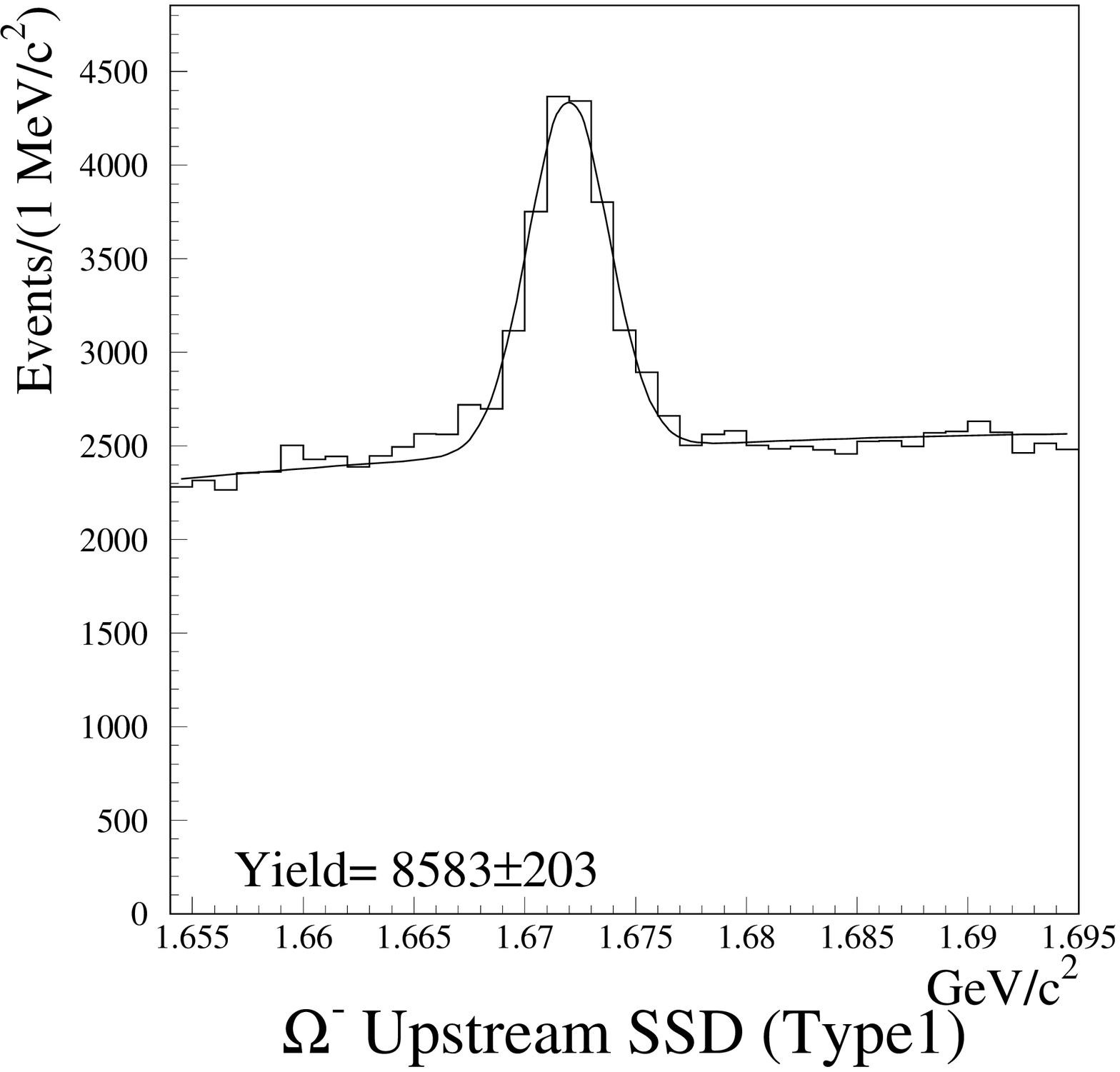}}
\end{center}
\caption{The invariant mass plots for the $\Lambda \pi^-$ 
and the $\Lambda K^-$ combinations for the category
where the decays occurs in front of the SSD detector. The plots
represent the full FOCUS data sample.}
\label{fig:castype1}
\end{figure}
 
\subsection{\bf Downstream Reconstructed $\Xi^-$'s and $\Omega^-$'s}

{\it Downstream} $\Xi^-$'s and $\Omega^-$'s are those which decay
downstream of the last microstrip plane and upstream of the
first MWPC plane.  They
are also referred to as `Type~2' decays.
A schematic of a typical $\Xi^-$ decay in this category
is presented in Fig. \ref{fig:cas2decay}. 
The decay distance for this category is more than three meters
along the beam direction. A very important advantage to these 
decays is that the $\Xi^-/\Omega^-$ leaves a track in the
microstrip detector before decaying. This track can in turn be
used for finding charmed particle baryon decay vertices.

\begin{figure}[htbp]
\begin{center}
{\includegraphics[width=7.5cm]{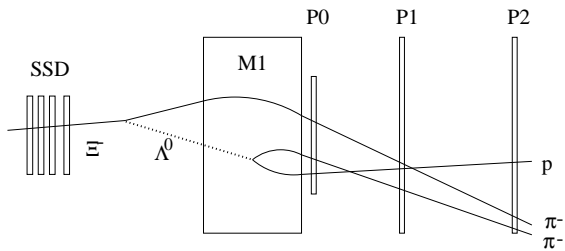}}
\end{center}
\caption{A schematic drawing in the bend view of the
spectrometer of a $\Xi^-$ decay which
occurs downstream of the silicon strip detector (SSD).
Only the front part of the spectrometer is displayed.}
\label{fig:cas2decay}
\end{figure}

The reconstruction algorithm begins by pairing each $\Lambda^0$
with every unlinked MWPC track in the event. An estimate is made
for the $\Lambda^0$-track vertex by computing the {\it z} intersection
of the $\Lambda^0$ vector and the track in the non-bend view
({\it xz}) plane. If this vertex is downstream of P0 or more than
50 cm upstream of the target, then the combination is rejected.
An estimate for the {\it x} and {\it y} positions of 
the vertex is given by
the {\it x} and {\it y} of the $\Lambda^0$ vector at the 
given {\it z} of the 
vertex (because the $\Lambda^0$ is neutral its path is not
deflected by the field of M1). The unlinked track is next traced
through the magnetic field of M1 to the position of the 
putative vertex. If the track is 3-chamber, then an iterative 
procedure is used whereby the track is assigned a different
momentum for each iteration until a good trace is made to the 
given vertex.  A better determination of the vertex {\it z} position
is made by computing the distance of closest approach
between the $\Lambda^0$ vector and track. (This iteration is
important because the track can pass through only one side of 
the magnet making fringe field corrections more significant.)

Next, the microstrip track of the candidate charged  
$\Xi^-/\Omega^-$ is found. The sum of the momentum vectors
of the $\Lambda^0$ and the track are used to form a 
candidate $\Xi^-/\Omega^-$. Unlinked microstrip tracks are
used and an attempt is made to match each one with 
the candidate $\Xi^-/\Omega^-$. If the $\Lambda^0$ under
consideration is a single-link Vee, then the Vee-linked microstrip
is used. Each candidate microstrip track is traced downstream
to the $\Lambda^0$-track vertex; if the vertex is within the
field of M1, then the magnetic trace is used otherwise the
microstrip vector is simply extrapolated to the vertex. 
Because the microstrip track has much better position resolution
than does the Vee, a better vertex position can now be
determined. The new {\it z} position is defined as the {\it z} 
where the
microstrip track and the MWPC track make their closest approach.
If the {\it z} position is downstream of P0 or upstream of the target,
the microstrip track is rejected as a candidate. Also, the 
$\Xi^-/\Omega^-$ vertex is required to be upstream of the 
$\Lambda^0$ vertex.

To remove spurious matches in this algorithm, a cut is made on
the candidate vertex. The candidate vertex is calculated
in two ways: the point where the $\Lambda^0$ and the MWPC track
make their closest approach, and the point where the microstrip 
track and the MWPC track make their closest approach. A cut
is made on the transverse distance (in the {\it x-y} plane) between
these two putative vertices. The second quantity on which a cut
is applied is the difference between the {\it x} and {\it y}
slopes of the
microstrip track and slopes given by the sum of the $\Lambda^0$
and the MWPC track momentum vectors. The {\it x} and {\it y} 
slopes must
agree to within 4 milliradians of the momentum vectors.

The MWPC track is assigned a pion mass for the $\Xi^-$
hypothesis and is assigned a kaon mass for the $\Omega^-$
hypothesis. For the $\Omega^-$ candidates, the MWPC track is
required to be identified by the \v{C}erenkov counters as being
consistent with a kaon hypothesis. 

Invariant
mass plots for downstream decays for the  $\Lambda \pi^-$ 
and $\Lambda K^-$ combinations
are presented in Fig. \ref{fig:castype2}. In this region one
finds that the $\Omega^-$ yield is a factor of 30 lower than the
$\Xi^-$ yield. The ratio of the $\Omega^-$ yield to the $\Xi^-$ yield
is reduced downstream of the SSD detector due to the shorter $\Omega^-$
lifetime relative to the $\Xi^-$.

\begin{figure}[htbp]
\begin{center}
{\includegraphics[width=6.5cm]{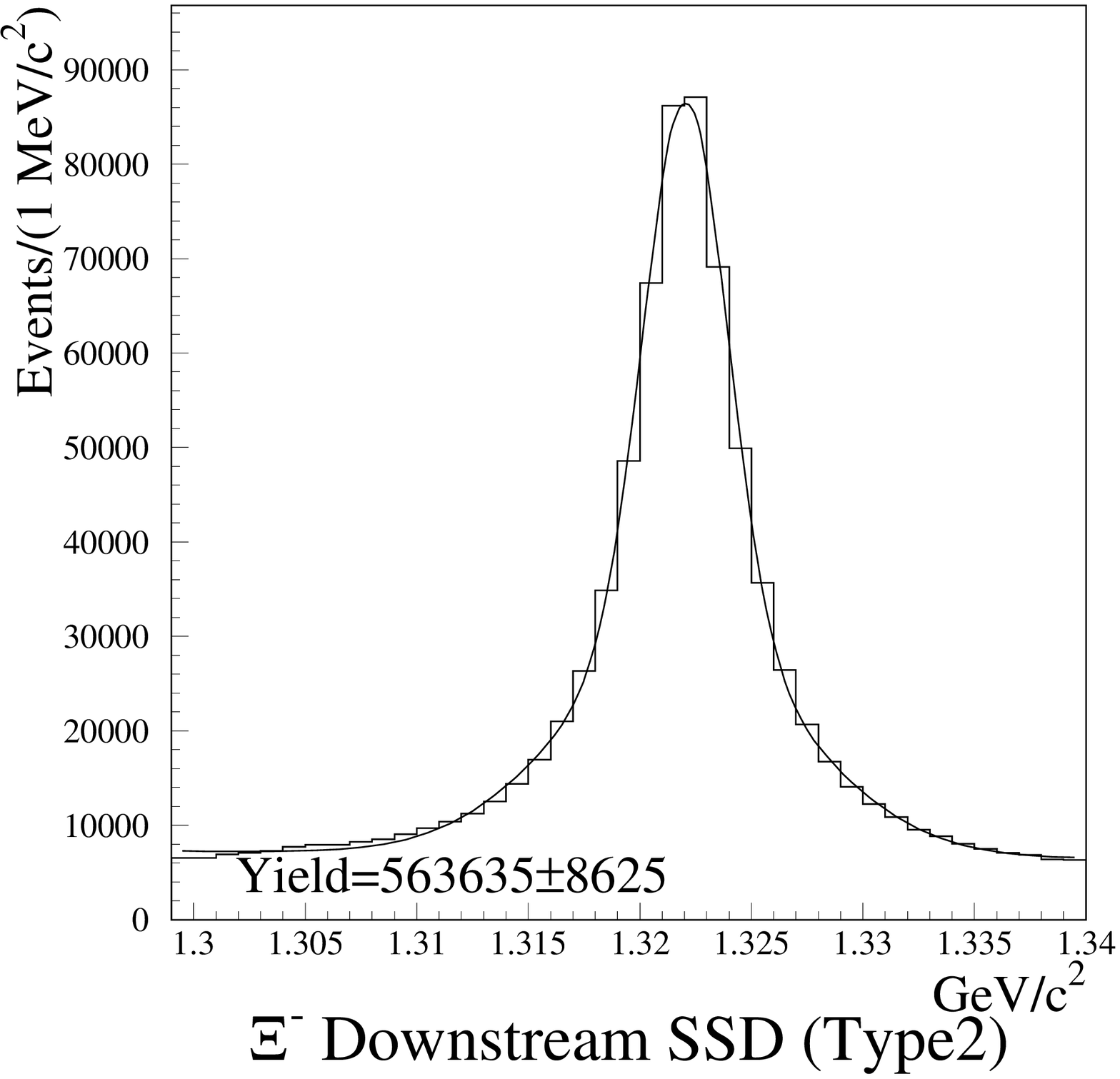}}
{\includegraphics[width=6.5cm]{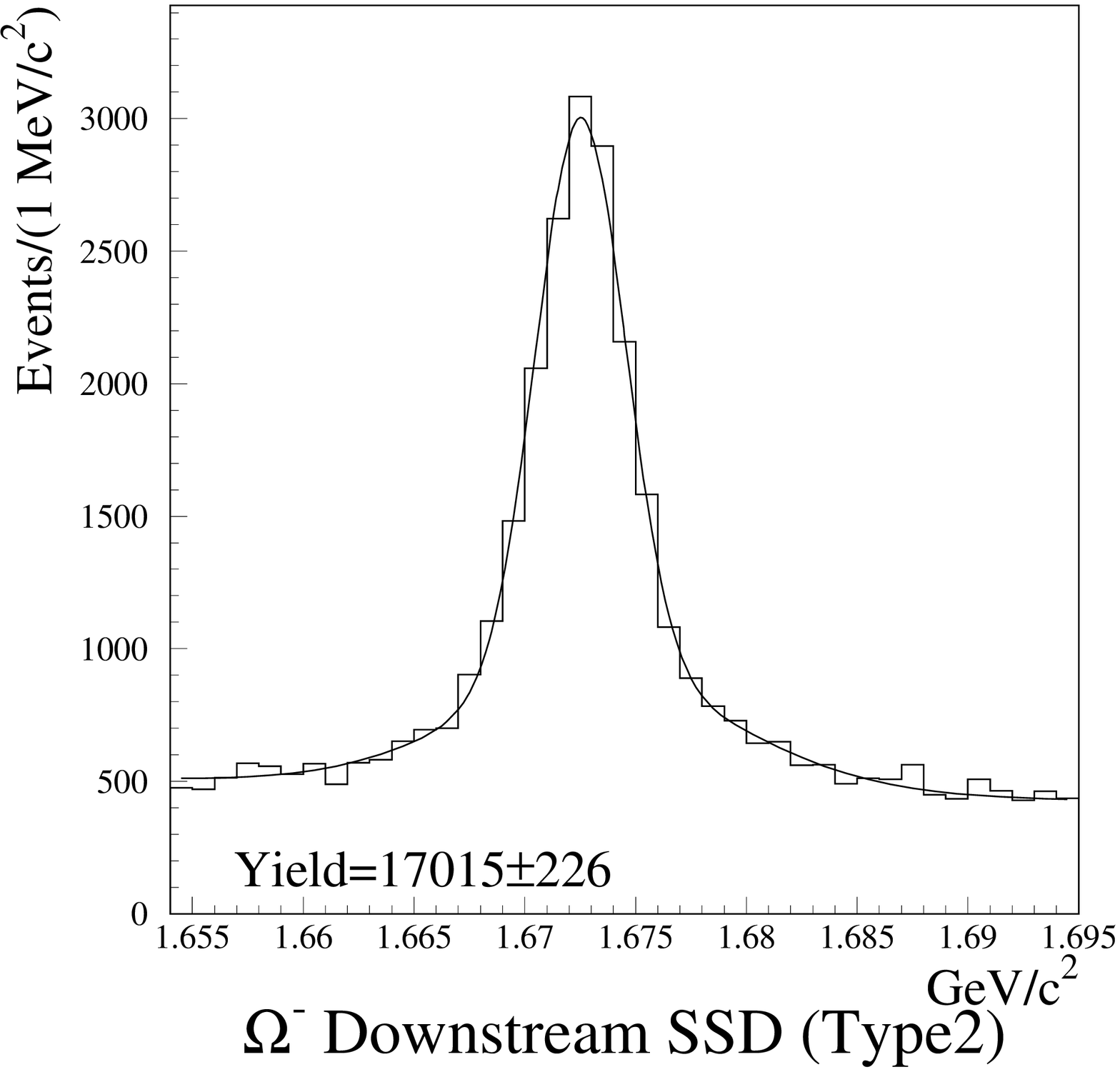}}
\end{center}
\caption{The invariant mass plots for the $\Lambda \pi^-$
and the $\Lambda K^-$ combinations for the category
where the decays occurs downstream of the SSD detector. The plots 
are for the full FOCUS data sample.}
\label{fig:castype2}
\end{figure}

\subsection{\bf Multivees}

A `multivee' is composed of three unlinked MWPC tracks and
one unlinked SSD track. A schematic drawing of a multivee decay
is presented in
Fig. \ref{fig:mvdecay}. While the category was designed to 
select three prong decays such as $K^-\rightarrow \pi^-\pi^-\pi^+$,
it has proven to be useful in recovering $\Xi^-$ and $\Omega^-$
decays where the $\Lambda^0$ decay vertex is close to the $\Xi^-$
decay vertex. It also works well in reconstructing Vees from 
$\Xi^-$ decays which 
open in the vertical plane of the magnet.

\begin{figure}[htbp]
\begin{center}
{\includegraphics[width=7.5cm]{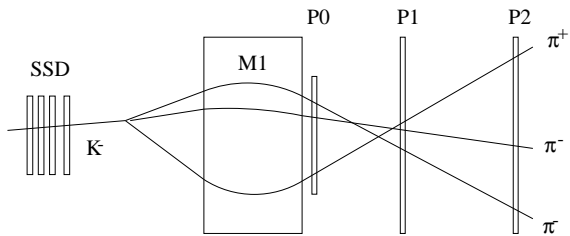}}
\end{center}
\caption{A schematic bend view drawing of the decay of
the $K^-\rightarrow \pi^+\pi^-\pi^-$ using three unlinked
MWPC tracks and one unlinked SSD track.}
\label{fig:mvdecay}
\end{figure}

Initially, three unlinked MWPC tracks are intersected in 
the {\it xz}
view and an unlinked SSD track with the closest distance of approach
at the {\it xz} vertex is selected as a match. 
The algorithm is separated 
into decays which occur upstream of M1  and within M1.

If the decay vertex
is upstream of the M1 magnet, then the SSD track is extrapolated
in {\it z} to the vertex and each of the stubs are traced to the
SSD {\it y} vertex position. If there are five chamber tracks, then they
are swum upstream and intersected with the SSD track to find a better
vertex position.   

If the decay occurs within the magnetic field, then there must be at least
one five chamber track. The vertex of the three unlinked MWPC tracks 
is found in the {\it xz} plane and the {\it y}
location of the vertex is determined by
swimming the tracks to the {\it z} location. If more than one
five chamber track exists, then the {\it z} location is determined 
using the 
combined {\it x} + {\it y} information. All remaining 
unlinked stubs are traced
to the {\it yz} vertex and their momentum is calculated. 
Next, the sum of the
three unlinked track momenta is found and assigned to the microstrip
track. This unlinked SSD track is now traced downstream to the 
{\it z} vertex.
Successful candidates must be near the vertex of the three unlinked
tracks  and must agree in slope to within 4 milliradians in both the 
{\it x} and {\it y} views to the combined momentum vectors of the three
tracks.

In Fig. \ref{fig:difmulti} the invariant mass distributions
for $\pi^+\pi^+\pi^-$, $p\pi^-\pi^-$, and $pK^-\pi^-$
combinations are presented. While the charge kaon decay 
events are not used in our analysis packages, the $\Xi^-$
decays and $\Omega^-$ decays are used. 

\begin{figure*}[htbp]
\begin{center}
{\includegraphics[width=4.5cm]{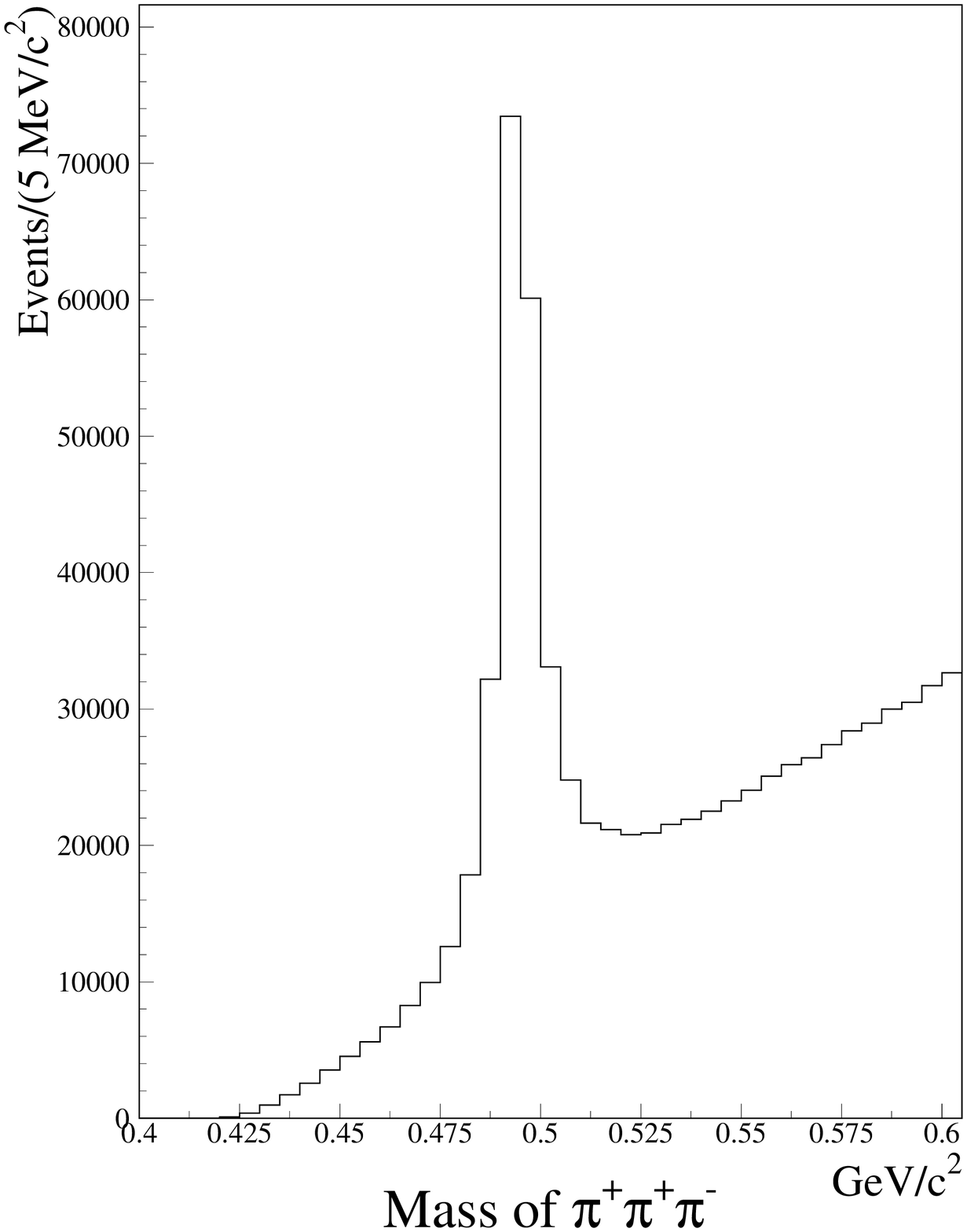}}
{\includegraphics[width=4.5cm]{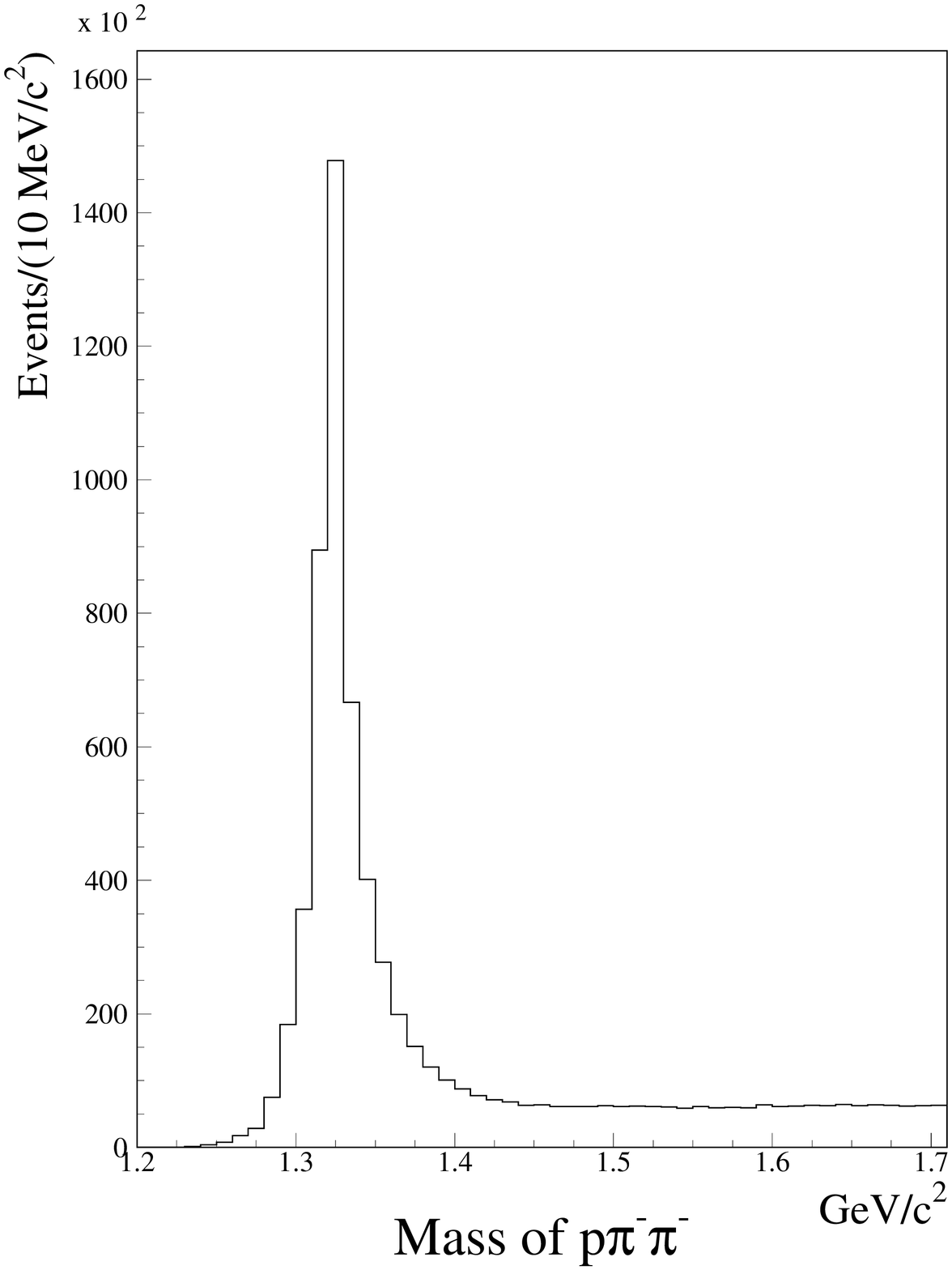}}
{\includegraphics[width=4.5cm]{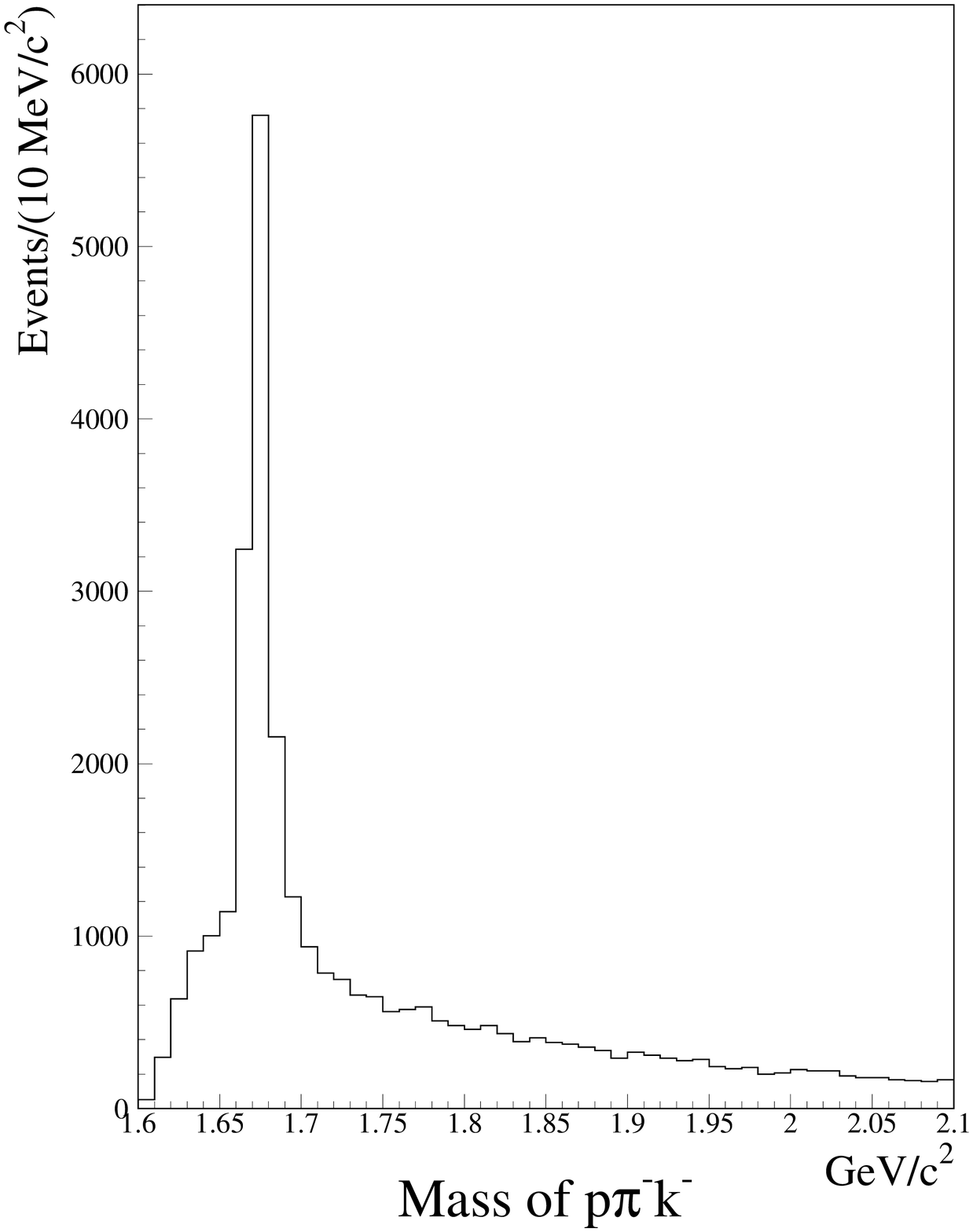}}
\end{center}
\caption{
Invariant mass distributions for $\pi^+\pi^+\pi^-$,
$p\pi^-\pi^-$, and $pK^-\pi^-$ combinations. Signals 
for $K^+\rightarrow \pi^+\pi^+\pi^-$, $\Xi^-\rightarrow \Lambda^0\pi^-$
where $\Lambda^0\rightarrow p\pi^-$,
and $\Omega^-\rightarrow \Lambda^0K^-$ where 
$\Lambda^0\rightarrow p\pi^-$ are clearly evident.}
\label{fig:difmulti}
\end{figure*}

\subsection{\bf Kinks}

The algorithm for reconstructing the $\Xi^-\rightarrow\Lambda^0\pi^-$
decays as presented in Fig. \ref{fig:knkdecay} is similar to
the $\Sigma$ decays discussed in Section 5. Unfortunately, there
is no good technique to reduce the sample with additional constraints
such as energy in the calorimeter or with \v{C}erenkov cuts.
For this reason we chose to only reconstruct events where the 
$\Xi^-$ decay occurred within the magnetic field of M1. There
are three advantages to using this subsample. First, the reconstructed
pion or kaon is a five chamber track and its momentum is well
defined by its passage through M2. Second, there is no two-fold
ambiguity in this category as the $\Xi^-$ also bends in the magnetic
field. Third, the decays are well-separated from any material 
and the background from large multiple scatters is significantly 
reduced. 

\begin{figure}[htbp]
\begin{center}
{\includegraphics[width=7.5cm]{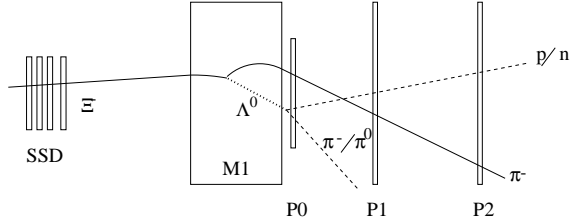}}
\end{center}
\caption{A schematic bend view drawing for a $\Xi^-$ decay inside
of M1 where the $\Lambda^0$ is not reconstructed.}
\label{fig:knkdecay}
\end{figure}

\subsection {\bf $\Xi^+_c$ Mass Plots}

Invariant mass plots from the full FOCUS data sample 
for the charmed particle decay $\Xi^+_c\rightarrow
\Xi^-\pi^+\pi^+$ for each of the four categories discussed 
above are presented in Fig. \ref{fig:cascpipi}. 
The mass plots were found with a significance of separation
cut of  $L/\sigma_L > 4$ between the secondary and primary
vertices.  $\Xi^-$'s 
are selected such that there is no overlap between categories.
Upstream $\Xi^-$'s or Type~1 decays have no overlap with the
other categories because they decay before the SSD detector.
Downstream $\Xi^-$'s or Type~2 decays are fully reconstructed
and are the cleanest of the decays. If a Type~2 decay occurs in
an event, the $\Xi^-$ Kink and multivee algorithms are not
run. The next cleanest category is the $\Xi^-$ multivees. If a 
$\Xi^-$ multivee is found, then the $\Xi^-$ Kink algorithm is
not used.  

\begin{figure*}[htbp]
\begin{center}
{\includegraphics[width=6.5cm]{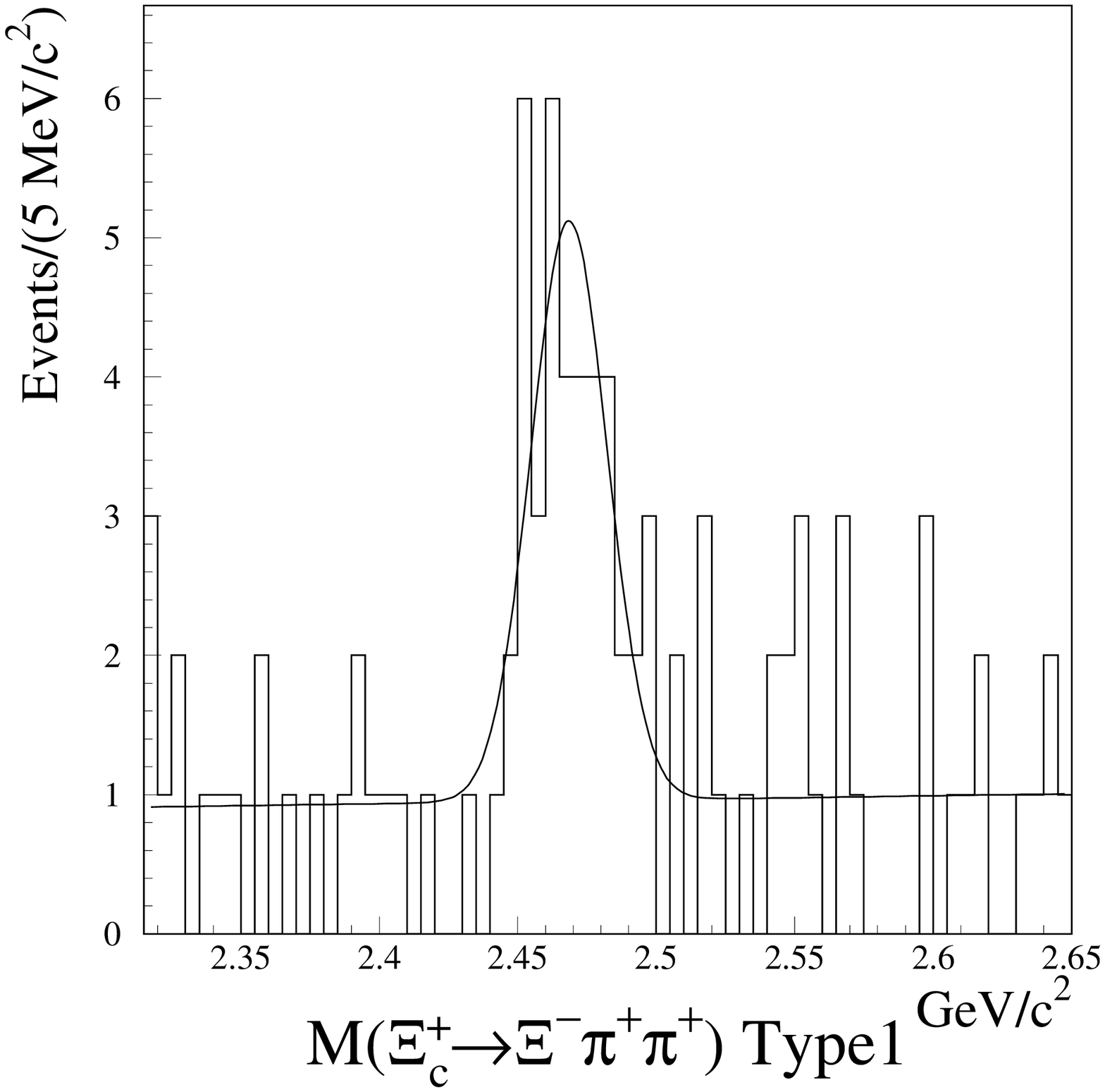}}
{\includegraphics[width=6.5cm]{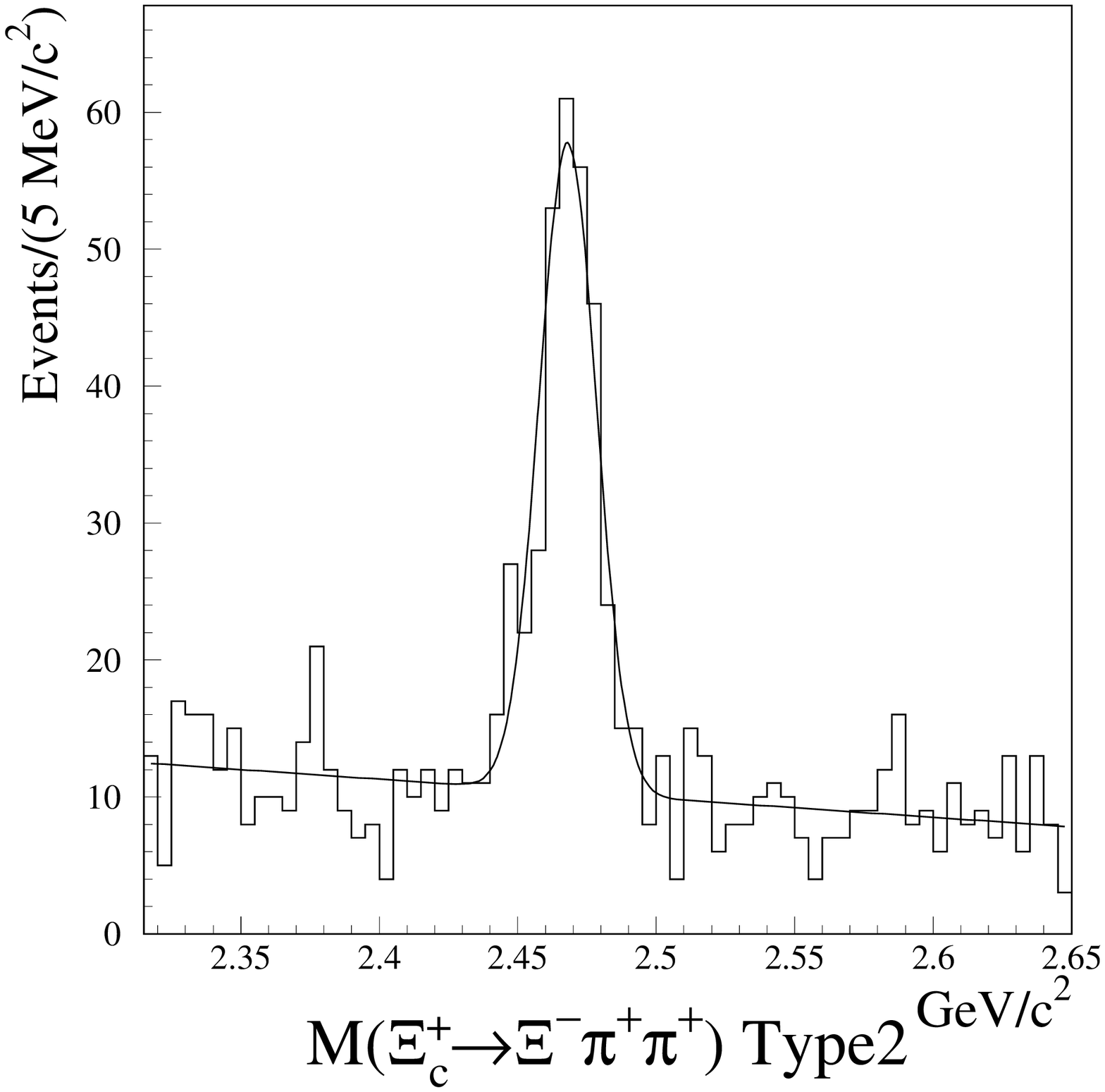}}
{\includegraphics[width=6.5cm]{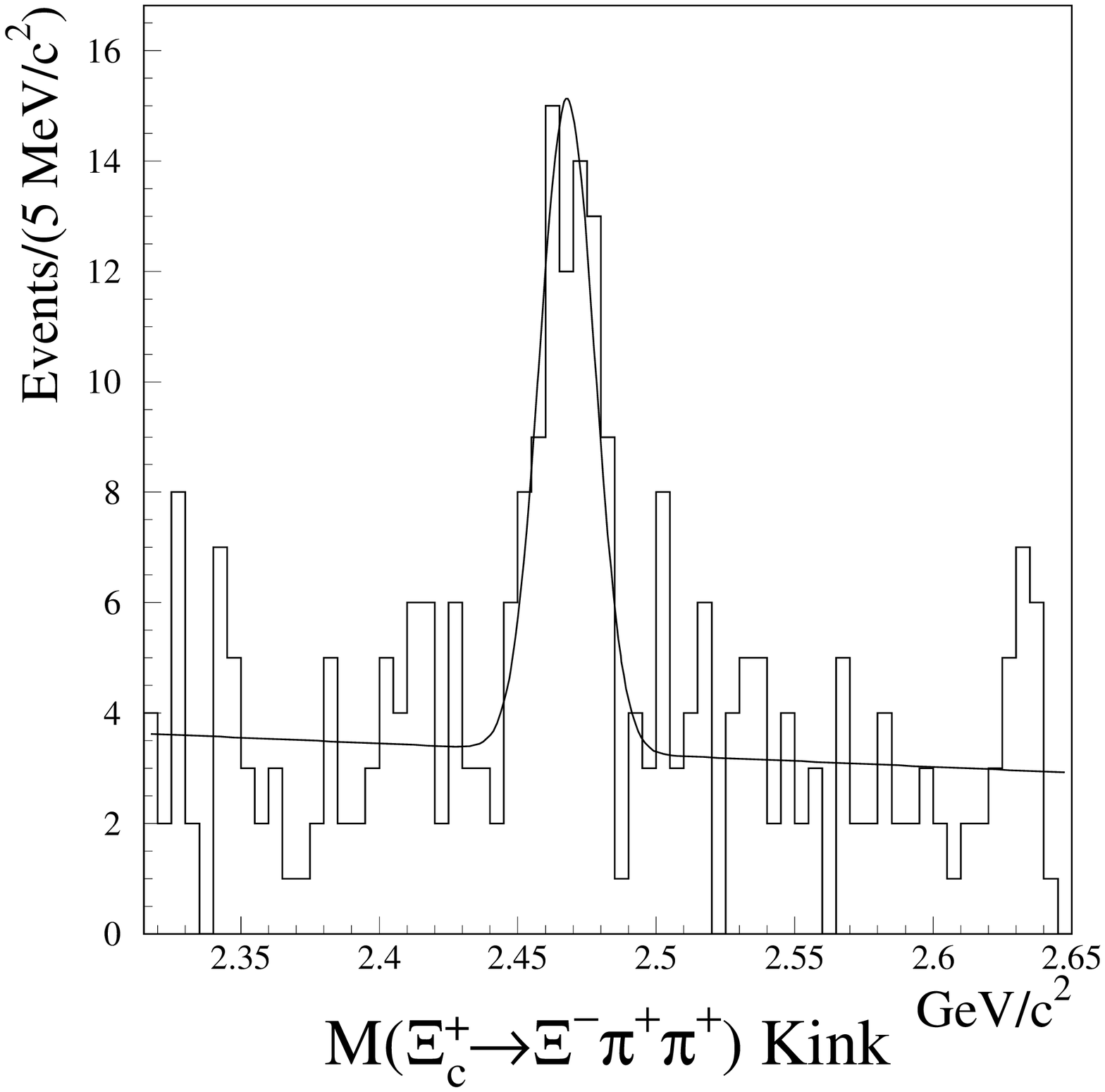}}
{\includegraphics[width=6.5cm]{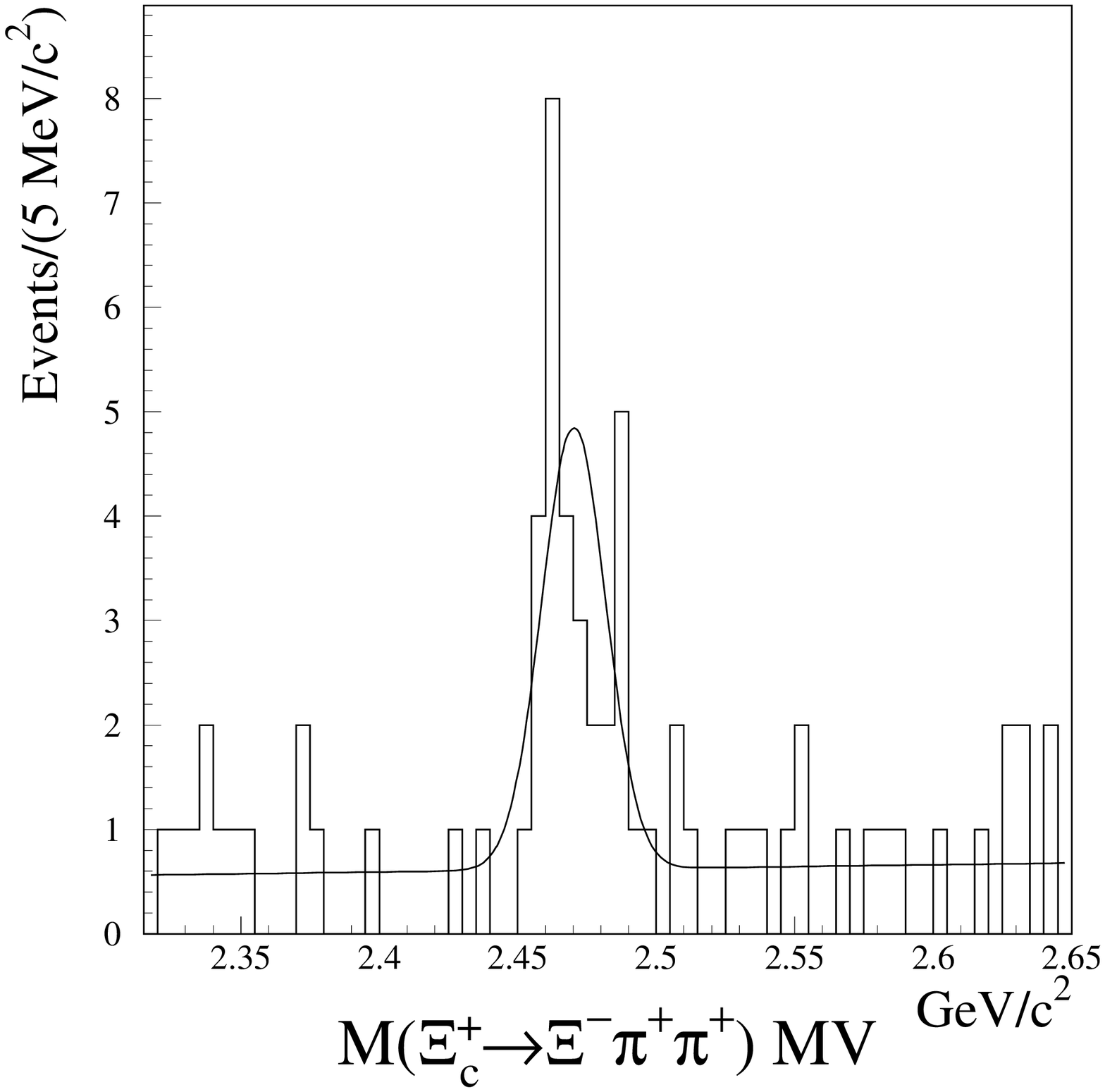}}
\end{center}
\caption{Invariant mass plots for the charmed particle
 decay $\Xi^+_c\rightarrow
\Xi^-\pi^+\pi^+$ using type1 (upstream~ $\Xi^-$) decays, using type 2 
(downstream~ $\Xi^-$) decays, using $\Xi^-$ Kink decays, and 
using multivee decays.}
\label{fig:cascpipi}
\end{figure*}
 
From an inspection of Fig. \ref{fig:cascpipi}, it is clear that 
the Type~2 category dominates the signal with 246$\pm$20 events, 
followed by the 
Kink category (59$\pm$9 events),  the Type~1 category (29$\pm$6 events), 
and the multivee category (25$\pm$4 events). 

\section {\bf Summary and Conclusions}
    
We have briefly described the tracking algorithms of FOCUS
and how the high resolution silicon microstrip system is 
integrated with multiwire proportional chambers. Further, 
we have described the various techniques developed
to reconstruct $K^0_S$ and $\Lambda^0$ decays in a multiparticle 
spectrometer. We have described the
two-fold ambiguity that occurs from $\Sigma^+$ and $\Sigma^-$ decays
to a single charged particle and an unobserved neutral decay. Through
these `Kink' kinematics we are able to observe charm baryon decays. 
Finally, we have combined the techniques of
Vees and Kinks and used these techniques in
the reconstruction of $\Xi^-$ and $\Omega^-$ decays. 
We believe the techniques decribed in this paper will prove useful 
to future
experiments and may serve
to show what is possible in multipurpose spectrometers.

\section*{Acknowledgments}

We wish to acknowledge the assistance of the staffs of Fermi
National Accelerator Laboratory, the INFN of Italy, and the physics
departments of the collaborating institutions. This research was 
supported in part by the U.~S.
National Science Foundation, the U.~S. Department of Energy, the
Italian Istituto Nazionale di Fisica Nucleare and Ministero
dell'Universit\`a e della Ricerca Scientifica e Tecnologica, 
the Brazilian Conselho Nacional de
Desenvolvimento Cient\'{\i}fico e Tecnol\'ogico, CONACyT-M\'exico,
the Korean Ministry of Education, and the Korean Science and 
Engineering Foundation.

\end{document}